# Computational Experiments: Past, Present and Future

Xiao Xue*, Xiang-Ning Yu, De-Yu Zhou, Xiao Wang, Zhang-Bin Zhou, Fei-Yue Wang

*Abstract:* **Powered by advanced information technology, more and more complex systems are exhibiting characteristics of the Cyber-Physical-Social Systems (CPSS). Understanding the mechanism of CPSS is essential to our ability to control their actions, reap their benefits and minimize their harms. In consideration of the cost, legal and institutional constraints on the study of CPSS in real world, computational experiments have emerged as a new method for quantitative analysis of CPSS. This paper outlines computational experiments from several key aspects, including origin, characteristics, methodological framework, key technologies, and some typical applications. Finally, this paper highlights some challenges of computational experiments to provide a roadmap for its rapid development and widespread application.**

*Index Terms*- Cyber-Physical-Social Systems; Computational experiments; artificial society; methodological framework; thought experiment; parallel optimization

## I. INTRODUCTION

There are two main purposes of system complexity research: law exploration and theoretical explanation. Here, the law can help us to clarify how system complexity occurs; the theory can help us to understand why system complexity occurs. The law is defined as the mapping between variables; the theory is defined as a causal explanation of the observed system complexity. However, powered by the rapid development of Internet, the penetration of the Internet of Things, the emergence of big data, and the rise of social media, more and more complex systems are exhibiting the characteristics of social, physical, and information fusion, which are called as Cyber-Physical-Social Systems (CPSS) [1,2]. Because CPSS involve human and social factors, the design, analysis, management, control, and integration of CPSS are facing unprecedented challenges. In scientific research, experimental methods often play an irreplaceable role, which not only promotes the establishment of the relationship between theory and facts but also drives the development of science and technology from the exploration and discovery of numerous experiments. Given this, experimental methods are being incorporated into CPSS research.

The traditional experimental method is generally a physical entity experiment, which cannot be carried out for the actual social system mentioned above. This is due to the following facts: **1)** The social complex system cannot be studied by the reduction method because the decomposed system would most likely lose the function of the original system, so it must be researched using the holism method [3]. **2)** This repeated test on the real system is not economically feasible due to the scale of social complex systems. **3)** Many complex systems that involve social governance are constrained by legislation and often cannot be tested or reshaped, such as national security, military preparedness, and emergency response, and so on; **4)** Social complex systems are highly human-related system, which invlove human in many scenarios. And, the testing of these systems may cause irreversible risks and losses, which are morally unacceptable [4].

Thanks for the support provided by National Key Research and Development Program of China (No. 2021YFF0900800), National Natural Science Foundation of China (No.61972276, No.61832014, No.62032016), and Shandong Key Laboratory of Intelligent Buildings Technology (No. SDIBT202001).

(Corresponding author: Xiao Xue; e-mail: jzxuexiao@tju.edu.cn)

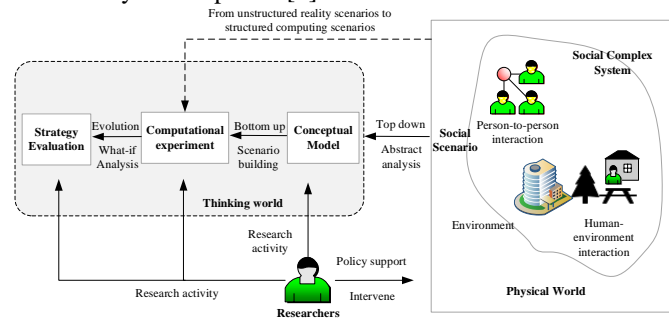

Fig.1 Schematic illustration of the computational experiments

Therefore, the research of social complex systems can only turn to "computational experiments", which can realize the quantitative analysis of complex systems by means of "Algorithmization" of "Counterfactuals" [5].Figure 1 shows the workflow of computational experiments. Firstly, this method abstracts the conceptual model of the reference system by constructing the autonomous individual model and their interaction rules from the micro-scale perspective. Then, by combining the complex system theory with computer simulation technology, the "digital twin" [6] of the real system can be cultivated in the information world. Furthre, by modifying the rules, parameters, and external intervention of the system, various experiments can be carried out repeatedly in a way of computation. Finally, According to the experimental results, the causal relationship between intervening variables and system emergence can be identified, which provides a new way to explain, illustrate, guide and reshape macro phenomena in reality.

Computational experiments are the natural extension and improvement of computer simulation. The difference between them lies in data-driven, emerging mechanisms, as well as multi-world interpretation and guidance theory. Besides the traditional description and prediction function of computer simulation, the computational experiment emphasizes the prescription function of the results, from "big law, small data" in Newton's era to "big data, small law" in Merton's era [7]. After more than ten years of development, the computational experiment method has become one of the mainstream methods to analyze complex systems. Besides multi-agent technology, its basic means include artificial systems and software-defined systems. Especially, the recent idea of the "digital twin" has been widely recognized in academic circles and even in society. It has become a new means of generating big data from small data and refining deep intelligence or exact knowledge from big data, which complements with artificial intelligence methods [8].

Compared with traditional experimental methods, the computational experiment method is characterized by the following three features: **1)** *Precise controllability*. By setting environmental parameters (such as geographical factors, agent distribution, etc.) and triggering events (such as time, location, type, scale, etc.), various scenarios can be accurately reproduced as the operating environment of the system. **2)** *Simple operation*. In the simulation process, it is easy to realize various extreme environments to evaluate the different performance indexes of the system, such as accuracy, response rate, etc. **3)** *Repeatability*. This advantage allows researchers to design different experimental scenarios and evaluate the effects of different factors (such as geographical environment, trigger event characteristics, etc.) on system performance [9].

As a scientific research methodology, computational experiments have been applied to some application studies that are risky, costly, or unable to conduct direct experiments in reality, including intelligent transportation system [10, 11], war simulation systems [12], socioeconomic systems [13], ecosystems [14], physiological/pathological systems [15, 16], political ecosystems [17], etc. Once an "artifitial laboratory" for complex systems has been established, extensive research on complex systems can be carried out in this laboratory. However, its development also faces a series of challenges, including the comparison and verification of computational models, the design method of computational experiments, the knowledge-driven and data-driven fusion, etc. The further study of these challenges will lay a solid theoretical foundation for the wider application of computational experiments. In addition, with the emergence of new technologies (such as reinforcement learning and digital twin), the computational experiment also needs to be integrated with new technologies to further enhance its application value in problem analysis and resolution.

The remaining parts of this paper are organized as follows. Section II introduces conceptual origins, application characteristics of computational experiment method, and our research motivation. Section III presents the methodological framework of computational experiments, which consists of five main links: modeling of artificial society, construction of an experimental system, design of experiments, analysis of experiments, and verification of experiments. Section IV provides three types of application cases: thought experiment, mechanism exploration, and parallel optimization. Section V discusses the problems and challenges that the computational experiment method may encounter in the future. Section VI concludes the paper.

## II. BACKGROUND AND MOTIVATION

Computational experiments are considered as a standard interdisciplinary research field. Its development history is closely related to the research of complex systems and the development of computer simulation technology. This section mainly explains the origin, characteristics, and our research motivation.

### A. Origin of computational experiments

With the deepening of scientific research, research objects of interest are becoming increasingly huge in scale and complex in function and structure. Thus, a discipline, known as the "Science of the 21st Century"—Complexity Science, came into being [18]. Complexity theory, as an interdisciplinary subject of a complex system, involves natural phenomena, engineering, economics, management, military, political and social fields; from the life phenomenon of a cell to the structure and mind of the brain; from the fluctuation of the stock market to the rise and fall of society, etc.

In terms of methods and paths, computational experiments have the potential to interrelate computer science, applied engineering, humanities, social sciences, and economic research, and provide new tools and means for researching complex systems. From the perspective of computer science, researchers of computational experiments extend the research scope to the social field and study numerous interesting and novel topics in collaboration with experts in other fields. Also, research depth is not limited to the description of objective things, but focuses on revealing the causes and evolution of objective things, and tries to predict their future development trajectories as accurately as possible. Fig. 2 shows the research of complex systems interlinks with the development of computer simulation technology to finally form the source of the computational experiment method.

The study of complex systems originated in the early 20$^{th}$ century. In 1928, Austrian biologist L.V.Bertalanffy first proposed the concept of "complexity" [19], whose thoughts originated from the evolutionism of British biologist Darwin C.R. and statistical physics of Austrian physicist Boltzmann L.E. Then, the research of complex system evolved, and underwent three stages:

**1)** 1950-1980: This stage focused on system science, which was represented by the old three theories (system theory [20], cybernetics [21], and information theory [22]) and the new three theories (dissipation structure theory [23], catastrophe theory [24], and synergetics [25]).

**2)** 1980-2000: This stage focused on the dynamics and adaptability of the system, which was represented by the self-organization theory (e.g. chaos theory [3, 26], fractal theory [27], critical theory [28]), and complex adaptive system theory [29].

**3)** 2000-now: This stage began to focus on the combination with data science theory, which is represented by the complex network [30, 31] and CPSS system [1, 2].

With the rapid development of complex system research, computer simulation technology as a tool has made great progress. The development of computer simulation technology makes it possible to map real systems in the information space. Traditional computer simulation holds the notion that the real system is the only one, and whether the simulation results are consistent with the real system is the only criterion to evaluate the experimental results. Based on this, the computational experiment takes the computer as the "artificial laboratory" to "cultivate" a possible macroscopic phenomenon in the real system and explore the law behind it. This provides a feasible way to analyze complex system behaviors and evaluate the intervention effects. The representative results of the computational simulation technology are as follows:

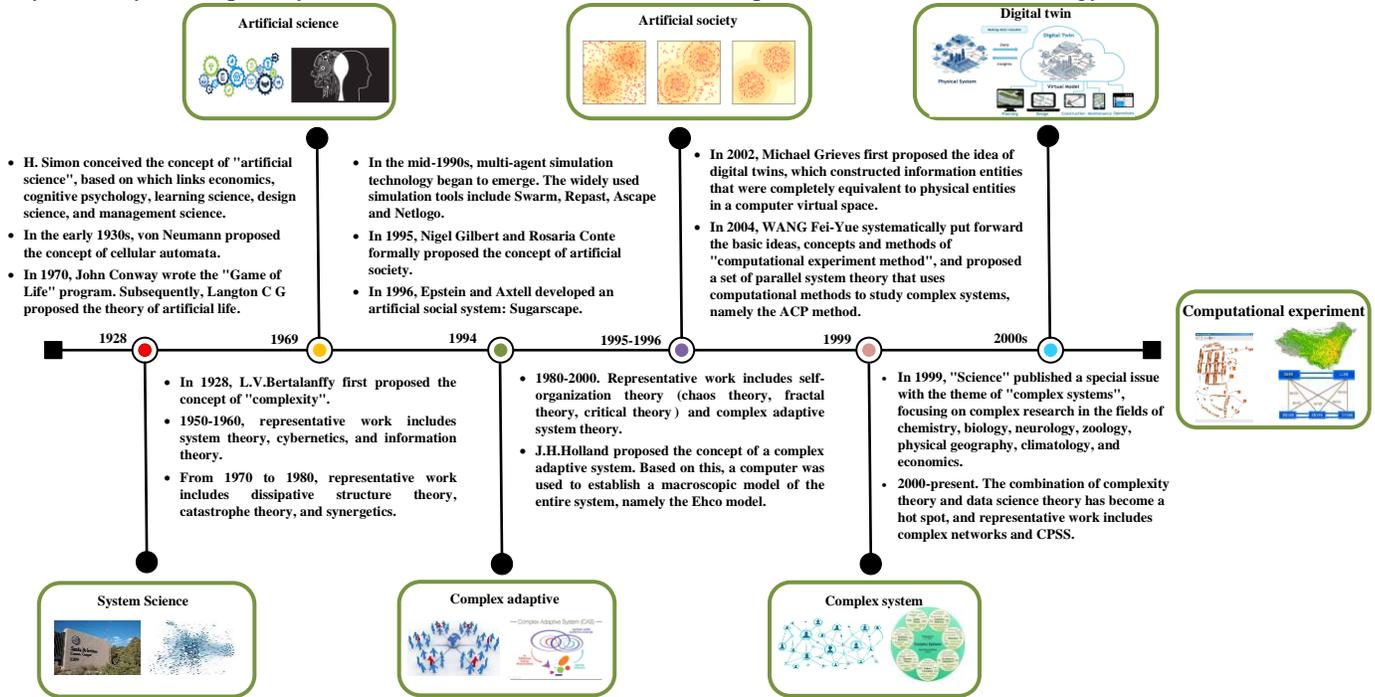

Fig.2 Conceptual sources of computational experiments

*1) Cellular automaton:* In the early 1930s, Von Neumann posited the concept of cellular automata (CA) [32]. The CA model emphasizes the interaction between autonomous individuals, focuses on the extraction strategy of entities in the system, and concerns the emergence attributes of the simple behavior of micro-individuals at the macro level. It aims to provide a general framework for the simulation of complex systems. The CA model first emphasizes entity cells and topology and then focuses on attributes/variables.

*2) Artificial life:* John Conway wrote the "life game" program [33] in 1970, which opened the prelude to artificial life research. Then, Langton C.G. proposed the artificial life theory [34], intending to construct a model system with the behavior characteristics of natural life systems by computer and other artificial media. Based on the concept of "chaotic edge", Langton, together with other scholars established various models to explore the evolution of artificial life, such as self-propagating cellular automata [35], Biods model [36], Ant colony model [37], "Amoeba World" [38], etc.

*3) Multi-Agent technology*: In the mid-1990s, multi-agent simulation technology began to rise and become an important means to study complex systems. The Agent in a multi-agent simulation represents an individual with a certain autonomy, intelligence, and adaptability. The main idea of multi-agent simulation is to simulate complex systems by modeling the interaction between a number of individual agents. The most widely used multi-agent simulation tools include Swarm [39], Repast [40], Ascape [41], and Netlogo [42]. With different advantages and disadvantages, these tools need to be selected according to specific needs.

*4) Artificial society:* In 1991, the Rand report first introduced the concept of "artificial society", where Agent technology was used to build social laboratories in computers and to test and evaluate different policies to ensure their effectiveness [43]. In 1995, Nigel Gilbert and Rosaria Conte published "*Artificial Societies - The Computer Simulation of Social Life*" [44], in which artificial society was formally proposed and became a relatively independent social science field. In 1998, the international academic journal *Journal of Artificial Society and Social Simulation* chaired by the University of Surrey was issued, marking the maturity of artificial society - Agent-based sociology simulation. Several classical artificial society models have emerged, including Epstein and Axtell's Sugarscape model [45], Arthur and Holland's artificial stock market model [46], the ASPEN model from the US Sandia National Laboratory [47], etc.

*5) Digital Twin:* Michael Grieves from the University of Michigan first proposed the idea of the "digital twin" in 2002. It hoped to build the digital twins equivalent to physical entities in the computer virtual space, establish the two-way dynamic feedback mechanism, analyze and optimize the physical entities [48]. In the industrial field, the digital twin is used to monitor, diagnose, predict, and optimize the operation and maintenance of physical assets, production systems, and manufacturing processes [49]. A digital twin can also be used to optimize the sustainable development of cities by capturing temporal and spatial effects. As a virtual replica of a particular city, digital twin technology allows city operators to develop different strategies. Digital twins have been implemented in some countries and cities, such as Singapore and Jaipur [50].

With the development of complex system science and computer simulation technology, Wang Fei-Yue formally proposed the concept of "computational experiment" in 2004, and formed the ACP method (Artificial Systems + Computational Experiments + Parallel Execution), emphasizing the circular feedback relationship between an artificial system and real system [51-55]. In recent years, with the rise of software definition [56], digital twins [6], reinforcement learning [57], service ecosystems [58], and other technologies, there is a growing connotation of computational experiment method [8]. As a bridge between the virtual world and the real world, computational experiments are providing new and effective computing theories and means to research complex systems.

*B. Characteristics of computational experiment*

Based on distributed thoughts and the bottom-up method, the computational experiment method can simulate the microscopic behavior of various entities in the real world with decentralized micro-intelligence models. By designing the interaction between individuals in experiments, the complex phenomena formed can reflect the macro law of system evolution. Through the well-designed computational model and simulation environment, computational experiments can be used as a powerful tool for reasoning, testing, and understanding of system complexity. Compared with the traditional analysis method, the computational experiment method can establish a variety of never-occurring virtual experimental scenes by changing the combination of internal and external parameters, and possibly several pressure experiments and extreme experiments. The role of different factors in system evolution is comprehensively, accurately, timely, and quantitatively analyzed. This will make it easier for researchers to explore the operation laws of complex systems and find effective interventions.

Computational experiment methods adopt the "multi-world" view of complex system research. When modeling complex systems, the degree of approximation to a real system is no longer the only criterion; however, the model is regarded as a "reality", which is a possible alternative form and another possible way to realize the real system. The real system is just one of the possible realities, with behavior "different" but "equivalent" to the model. Table 1 compares the characteristics of computer simulation, digital twin, and computational experiments. In short, the computational experiment system is a software definition of the real system, which is not only the digital "simulation" of the real system but also the alternative version of the real system (or other possible situations). It can provide efficient, reliable, and applicable scientific decisions and guidance for the design, analysis, management, control, and synthesis of real complex systems.

Table 1 Comparison of features between computational experiments and similar concepts

|  | **Computer Simulation** | **Digital Twin** | **Computational experiment** |
|---|---|---|---|
| Research object | Physical systems with well-defined and specific structures, which emphasize the modeling and simulation of independent unit. | Cyber-physical system, which covers the modeling and simulation of the entire integration process, from design, manufacturing, operation, to maintenance. | Social complex systems, which emphasize the fusion and interaction of social space, information space, and physical space. |
| Research means | Based on the principle of similarity, a mathematical model is established by top-down decomposition, which has the homomorphic relations as the actual or envisaged system. The computer can output the same results as the physical system by numerical calculation. | It can be used to solve non-linear and uncertain problems that cannot be solved by the traditional mechanism model. Furthermore, it can form an evolving system with machine learning. There are two optimization modes: model-driven and data-driven. | The computational experiment can simulate and deduce scenarios that have never occurred in real world. It is not required that the computational model must completely reproduce the behavior of the actual systems. |
| Research objectives | It focuses on the accuracy of modeling, that is, whether it can accurately reflect the characteristics and states of physical objects. Thus, it can guide the design and optimization of actual systems. | It focuses on how to reflect the data interaction between digital objects and physical objects, as well as the dynamic changes of the system. Thus, the digital twin has the value of continuous improvement in industrial applications. | It emphasizes that the experimental system is not only a digital simulation but also an alternative version for real system. The experimental results can provide decision support for the research of complex systems. |
| Application fields | It has been applied in various fields, including transportation, aerospace, industrial manufacturing, meteorological forecasting, electronic information industry, and so on. | It can lay a solid foundation for enterprise digitalization, which can be applied in innovation validation, virtual debugging, data monitoring, remote diagnosis, remote maintenance, etc. | It has led to the emergence of multiple interdisciplinary research fields, such as computational sociology, computational economics, computational finance, computational epidemiology, etc. |
| Limitations | Due to the lack of sufficient theory and prior knowledge, it is difficult to use top-down modeling to accurately describe and analyze complex systems. | Due to the lack of consideration of "human factors", it is unable to explain and analyze complex social phenomena. | No consensus is reached on how to prove the validity and equivalence of the computational model. It is easy to be questioned whether the experiment can reflect reality. |

A wide range of the application of computational experiments exists. In order to obtain the expected research goal, it is necessary to balance reality and abstraction when constructing a computational model. For example, highly realistic models may have significant policy value with little or no theoretical value; conversely, highly abstract models may provide profound scientific insights, but they can only provide results that are rarely directly applied in terms of policy practice. According to the application characteristics of a computational experiment, related applications can be summarized into the following four categories:

*1) Highly abstract*

The model has only a few qualitative similarities with the reference system and does not attempt to replicate any quantitative features. These models are mainly used for the theoretical analysis of basic science, not operational strategy analysis. Some early social simulation models belong to this group, such as the heatbug model [59], the boids model [36], and so on.

*2) Moderately abstract*

The model can exhibit convincing qualitative characteristics and meet some quantitative criteria. Although these models are largely theoretical, they can provide several applicable insights with valuable impacts on policy development. For example, although the classical Schelling model [60] is quite abstract, it reveals important insights into social isolation phenomena.

*3) Moderately realistic*

Although the model belongs to the qualitative category, it accords with the quantitative demands in terms of important characteristics. Experiment-based social computing studies are most interested in such models. For example, due to public safety considerations, it is impossible to carry out some destructive social experiments in the real environment, such as some public safety events. Many extreme pressure experiments can be simulated without risk by changing the experimental conditions and setting different variables.

*4) Highly realistic*

In terms of quantitative and qualitative characteristics, the experimental output of this type of model is the most consistent with empirical data. The highly realistic simulation can be compared from multiple dimensions to the reference system, including spatial features, temporal features, or organizational patterns. Such models are widely used in business and government organizations. Public policy involves highly uncertain areas (such as social networks and human behavior), resulting in complexity in its analysis, formulation, and implementation. Taking artificial society as an alternative version of the real system, the computational experiment methods can be used to analyze the effect of policies (such as economic stimulus, laws, and regulations, etc.), to improve the scientific level of public policy-making.

*C. Research motivation*

As researchers pay growing attention to the computational experiment method, it has spawned many inter-disciplinary research fields, such as computational economics [13], computational finance [61], computational histology [62], computational epidemiology [63], etc. In February 2009, 15 top scholars from global renowned universities (Harvard University, Massachusetts Institute of Technology, etc.) published the paper "The Age of Computational Social Sciences" [64] in *Science*. In 2012, 14 famous European and American scholars jointly issued "*Declaration on Computational Social Science*". Today, "Computational Social Science" has become the name accepted by mainstream academia [65]. In 2020, these scholars published a joint paper in *Science*, emphasizing the problems and challenges in the development of computational social science [66].

Note that, research on computational experiments requires multi-disciplinary knowledge, such as computer science, social science, systems science, artificial intelligence, computer simulation, and many other disciplines. Despite the efforts of researchers, a complete and advanced theoretical system has not been formed, and a chasm still exists between its theoretical development and practical application. In order to promote the development of this field, the state-of-the art of computational experiments is reviewed in this paper. It is hoped to help readers construct a complete knowledge system of computational experiment methods and lay a solid foundation for its subsequent development. We focus on the following three aspects:

*1) What is the technical framework for computational experiments?*

As a multi-disciplinary field, the computational experiment method involves a growing spectrum of knowledge. When beginners first interact with computational experiments, it is easy to drown in the vast literature, and also difficult to clarify the relationship between computational experiment methods and various techniques. This dilemma is exacerbated by the emergence of new technologies and applications. The resulting problem is whether a unified and general methodological framework could be developed to guide the application of existing computational models or the development of new computational models, thus greatly reducing the difficulty of implementing new applications. Based on this, the third section represents a general methodological framework to illustrate the related key technologies of computational experiments.

*2) How to implement a specific application of computational experiments*

Because of the diversity and uncertainty in the application field, as well as the subjectivity in the construction of artificial societies, there are always different opinions in the academic community on whether the experimental system can represent the original system. High confidence is the basis for applying the computational experiment methods. In addition to the breakthrough in experimental design and model verification technology, it is necessary to analyze successful application cases and summarize the valuable experience. Only in this way can the computational experiment method learn from other methods and transform into a powerful tool to understand the operation laws of complex systems. Therefore, section IV reviews typical computational experiment cases from three levels: thought experiment, mechanism exploration, and parallel optimization.

*3) What is the future development trend of computational experiments?*

The computational model is the core of the computational experiment method and the repository for applying different areas of knowledge. In recent years, endless waves of new technologies have emerged, such as digital twins [6], generative adversarial networks (GAN) [67], reinforcement learning [57], etc. These technologies have a considerable impact on the construction of computational models. Based on this, how to improve the computational model design using various new technologies has been a key challenge in the sustainable development of computational experiment methods. Accordingly, section V focuses on three aspects: **(1)** how to define the artificial society using big data, i.e. describing intelligence; **(2)** how to predict the future using computational experiments, i.e. predictive intelligence; 3) how to adapt the feedback intervention in the real world, i.e. guiding intelligence.

## III. METHODOLOGICAL FRAMEWORK FOR COMPUTATIONAL EXPERIMENTS

In Fig.3, the methodological framework of computational experiments can be summarized as a five-step feedback loop: modeling of artificial society, construction of an experimental system, design of experiments, analysis of experimental results, and verification of experimental models. Based on the framework, this section will elaborate on the key technique used in each step of the computational experiment method.

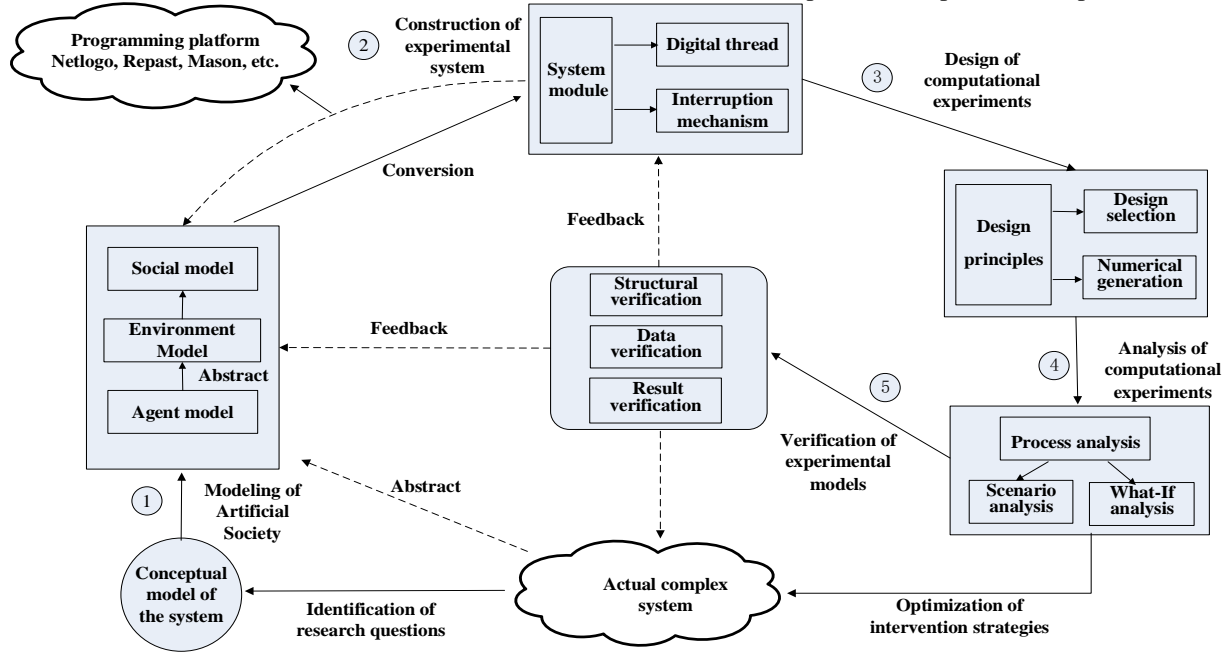

Fig. 3 Methodological framework for computational experiments

### A. Modeling of artificial society

Artificial society is a method of simulating human society in computers [68]. Compared with the general simulation model, the artificial society model can describe more complex systems. Here, uncertainty in individual behavior, as well as a complex interaction between individuals, exists. After defining the structure, elements, and attributes of artificial society, it is necessary to map the complex system into a multi-agent system in information space, which focuses on the individual model, environmental model, and social model. The individual model is an adaptive behavior mechanism that describes the individual agent; the environmental model is a description of the social attributes of the individual agent; the social model is a description of the mechanism of system evolution. The details are as follows:

#### 1) Individual model

An agent in the artificial society is an individual with independent ability, which corresponds to the biological individual or biological group in real society. The individual agent model is a container for applying multi-disciplinary knowledge, which can be customized according to application problems, including agent structure, learning ability, interactive mechanism, etc. Individual agents in artificial societies can adopt homogeneous structures or heterogeneous structures. As shown in Fig.4, the typical structure of an individual Agent consists of four parts: perception, decision, reaction, and optimization [69]. The information control flow in the Agent structure connects the parts into a unit. The following is a formal expression of the Agent structure, which is described by a set of properties related to time t.

$$Agent = <R, S_t, E_t, Y_t, V_t, N> \quad (1)$$

Wherein, $R$ represents the features that Agent does not change with time, such as identification; $S_t$ represents the feature that Agent changes with time, such as the role of Agent; $E_t$ is a collection of external events that Agent perceives and stimulates their state and behavior; $Y_t$ is the decision-making mechanism adopted by Agent in the process of feeling external stimuli and interacting with other agents; $V_t$ is a collection of Agent behaviors, including all behaviors taken spontaneously and stimulated by external events; $N$ is the constraints to which Agent is bound, including the environment, other agents, and the mission objectives.

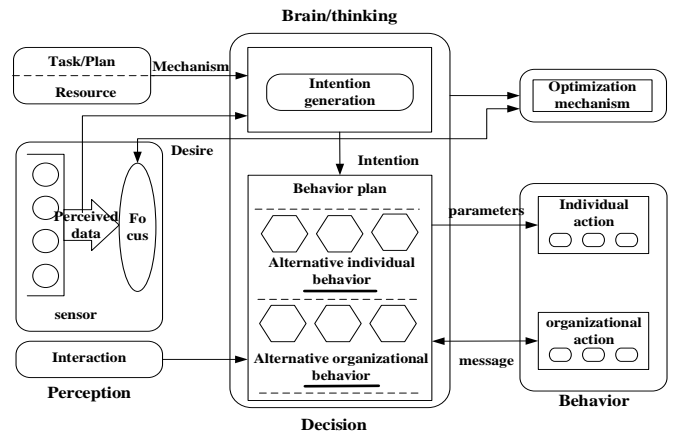

Fig. 4 Structure model of individual Agent

In the artificial society, the individual Agent is neither unconscious nor lacks initiative. The learning process of each agent is an important driving force for the evolution of the system. In the operation process, the Agent will interact with the environment to experience problem solving, update the rule base, and impact the decision-making mechanism. According to the strength of individual consciousness (or rationality), individual learning models can be classified into three categories: **(1)** non-conscious learning, including reinforcement learning [70] and parameterized learning automaton [71]; **(2)** learning by imitation [72]; **(3)** belief-based learning, including fictitious play [73], random dynamic learning [74], Bayesian rational learning [75], etc.

*2) Environmental model*

In the computational experiment, the environment model is the mapping of the actual physical environment in the computer, which is the activity place that the Agent relies on. According to the modeling method, the artificial society system can be divided into physical modeling and grid modeling. The physical artificial society is to abstract various environmental elements of real society (such as buildings, road traffic, and climate conditions) into physical models. At present, many typical artificial society systems are implemented in this way, such as EpiSimS[76], etc. The grid artificial society does not focus on specific environmental elements, but on modeling environmental space. Discrete grids are used to describe the existence of physical space and the properties of the environment, such as the Sugarscape model [45].

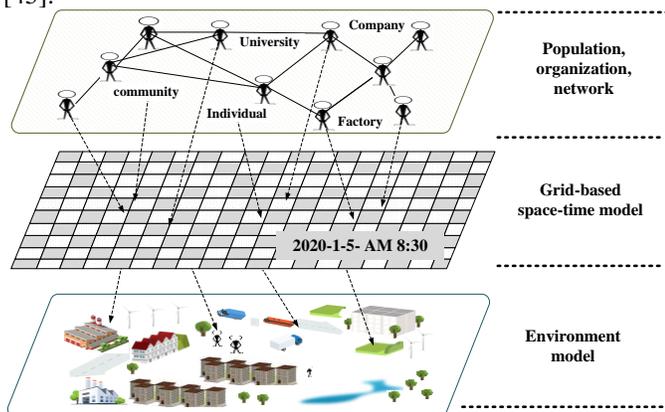

Fig. 5 Abstract hierarchy of environmental models

Because of the different sizes of artificial society, the model granularity of environment elements is different. The granularity of the environmental model is relatively coarse in the large-scale artificial society scene. For example, in the analysis of the global or urban-level spread of disease, the traffic networks, such as aviation networks, are generally considered as a simple abstract network. In the small-scale artificial society, the small granularity model may be established for the natural environment, including buildings and transportation. For larger artificial society scenes, GIS is usually used to build geospatial models. For small-scale artificial society scenes, visual scenes can be established by 2D or 3D display technology. At the same time, the coordinate system is established to determine the position in geographical space (Fig. 5).

Because of the limitation of actual conditions, the initial setting of the environmental model can only be statistical characteristic data. Therefore, it is necessary to study the generation algorithm of artificial society initialization data, including statistical characteristics of Agent (total number, sex ratio, age distribution, etc.), geographical distribution of Agent, demographic and social relation attributes, statistical characteristics of environmental entities (total number, type, population accommodated, etc.), the geographical distribution of environmental entities, etc. It is the basic idea of environmental modeling to reconstruct the specific characteristics of each individual in the group from the statistical feature of population data. In the process, two aspects of consistency need to be met: **(1)** the number of Agents generated is statistically consistent with the real world; **(2)** the internal logical structure (i.e. the asoociated relation) of generated Agent is consistent with the real world, i.e. the associated relation of Agents is consistent with the real world.

*3) Social model*

The social model describes the cyclical mechanism of artificial society, including "interaction" rules between Agents, environments, Agent and environment. These rules can be either a mapping of real social rules or an artificial hypothetical rule. As shown in Fig. 6, the evolution order of an artificial society is defined from three levels using the bottom-up framework. The bottom layer is the individual evolution space, which is used to simulate the phenomenon of genetic evolution in social systems; the middle layer is the organization evolution space, in which the individuals can enhance their ability through imitation and observation; the top layer is the social evolution space, which is used to simulate the phenomenon of accelerated evolution of society promoted by social influence. The social influence established by drawing excellent knowledge from the bottom layer can guide the evolution of the bottom individuals [58].

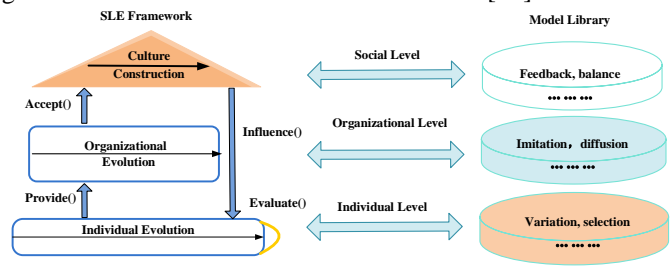

Fig.6 The social molding framework

This analytical framework is also characterized by the complex system theory, given the nesting and causality of evolution mechanisms between layers. These three layers and their interactions constitute a complete and abstract circular analysis structure. Each layer focuses on the different dimensions of the artificial society evolution process, including individuals, organizations, clusters, and countries. Relevant modeling techniques can be selected according to specific requirements. The implementation logic of SLE is shown in Table 2.

TABLE 2 The implementation logic of SLE

| |
|---|
| Definition-Agent: Used to represent individuals in an artificial society. |
| **Step 1: Individual evolution** <br> Variation is the generating mechanism of diversity and the source of social system evolution. If there is no mutation and innovation, there is no evolution. Agent's variation rules can be set according to various evolutionary algorithms in the model framework. Some are designed to imitate the evolutionary functions of biological systems, such as the artificial neural network (ANN) [77], genetic algorithm (GC) [78], evolutionary strategy (ES) [79], etc. |
| **Step 2: Organizational evolution** <br> The selection mechanism is a mechanism of decreasing diversity. It evaluates the adaptability of individuals by some criteria, selects the evolution units with high adaptability, and eliminates the evolution units with low adaptability. In the model framework, the evolution rules of the |

organization can be set according to the evolution characteristics of the biological community, such as ant colony optimization algorithm (ACO) [37], particle swarm optimization (PSO) [80], artificial bee colony algorithm (ABC) [81], etc.

**Step 3: Social emergence**

After fierce competition, some elites will emerge from the group. Other individuals can improve their ability to survive in the ecosystem by imitating and learning their behaviors. This stage is the social evolution stage. There are three typical diffusion and evolution models: contagion model [82], social threshold model [83], and social learning model [84].

**Step 4: Second order emergence**

Social space, in turn, acts on individual space, by which macro phenomena can affect micro-individuals. In order to simulate the phenomenon that culture can accelerate the evolution of individuals, we design feedback rules in the model framework. The selection mechanism of intervention strategy affects the variation level of an individual and the whole system.

**Step 5: Next cycle**

Over time, some elites may fall behind, and some new individuals with stronger capabilities will become new elites. Finally, the evolutionary equilibrium of the whole system breaks and transitions into the next cycle.

*B. Construction of an experimental system*

By using the bottom-up method, the computational experiment system can simulate the whole complex system with a decentralized micro-intelligence model. First of all, it is necessary to simulate the microscopic behavior of intelligent entities, to realize the complex phenomenon formed by the interaction of simple elements, and then to explore the macroscopic law of the whole system through the experiment design. The computational experiment system focuses on how to integrate various models with a digital thread to support the implementation of artificial society and the identification of intervention strategies.

*1) System module*

There are two development methods of the artificial society model: the first is self-programming to achieve greater freedom of modeling; the second is the adoption of specific platforms to achieve greater development efficiency. At present, the development of the artificial society model is still in its infancy. The importance of development efficiency is higher than that of modeling freedom. Generally, researchers generate some framework codes with a development platform and manually prepare the codes of specific functions to reduce the programming workload. If the platform is trusted, the code it automatically generates is also trusted. At present, it has been a mainstream trend to develop artificial society models using platforms, such as Swarm[39], Repast[40], Mason[17], Netlogo[42], etc.

As shown in Fig.7, the development architecture of the artificial society can be divided into three modules: preprocessing, plan execution, and output. In the preprocessing module, the status value of agents, the external status value, and instructions provided by the virtual environment are collected at the same time. In the plan execution module, the system selects the knowledge rules in the knowledge base and the model in the model base, calculates and executes the behavior of various Agents in the simulation environment, and outputs the calculation results to the behavior selector to set the Agent sub-target in the next moment. Based on behavior selection results and external environment data, the Agent behavior learning rules are used to configure Agent behavior, and the behavior selection parameters can be modified by feedbacks. Finally, the internal state value of Agent is updated, and the behavior state information is passed to the virtual environment in the simulation module. In the output module, the dynamic update of the virtual environment and agents is realized to enter the loop calculation at the next moment.

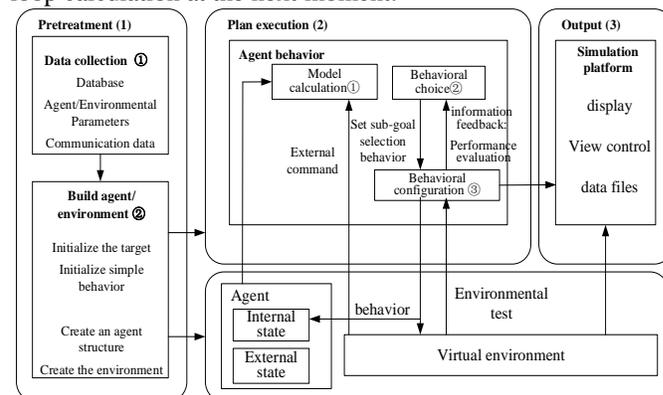

Fig. 7 System module of computational experiment system

*2) Digital thread*

It is important to realize the uplink/downlink data interaction between the computational experiment system and the real world. Only in this way can the computational experiment have the application value of continuous improvement. A large amount of available data is gradually becoming the original driving force for computational experiments. It plays an important role both in the process of model abstraction & simplification and in the process of model correction & verification. Based on this, we introduce the concept of "digital thread"[85] in the industrial Internet to describe the whole life cycle process of the computational experiment method, which can provide the virtual representation of key elements (digital twin) and the correlation between them. This can be used to trace the experimental phenomena forward or backward, thus assisting strategy evaluation, impact analysis, defect backtracking, and so on. From the perspective of knowledge engineering, the construction of artificial society is to organize and encode the knowledge of the real system so that it can accept and process the data input in the real world.

The construction of artificial societies is a complex multi-stage, multi-factor, and multi-product process. From the top-down perspective, artificial society modeling includes a conceptual model, a mathematical logic model, and a simulation model, which have mutual dependence and constraint relationships. In the concept layer, the system is modeled according to physical concepts. System state variables with physical meaning are described by physical quantities, such as force, velocity, power, etc. The modeling of the logical layer is closely related to the concept layer, which mainly builds the relationship between various state variables. In a continuous system, the relation is usually represented by a partial differential equation or ordinary differential equation. These equations caonnot be applied to complex systems, so they need to be converted into numerical models which could be adoptable. In the simulation layer, after completing the transformation process from the upper model (logical layer) to the lower model (numerical calculation model), it can be considered that the computational experiment system can have the capability to support the calculation of the conceptual model and the logical model. As shown in Fig.8, the digital thread can be seen as a bridge connecting different models, which aims to show the system evolution history and special state transition during the experimental cycle.

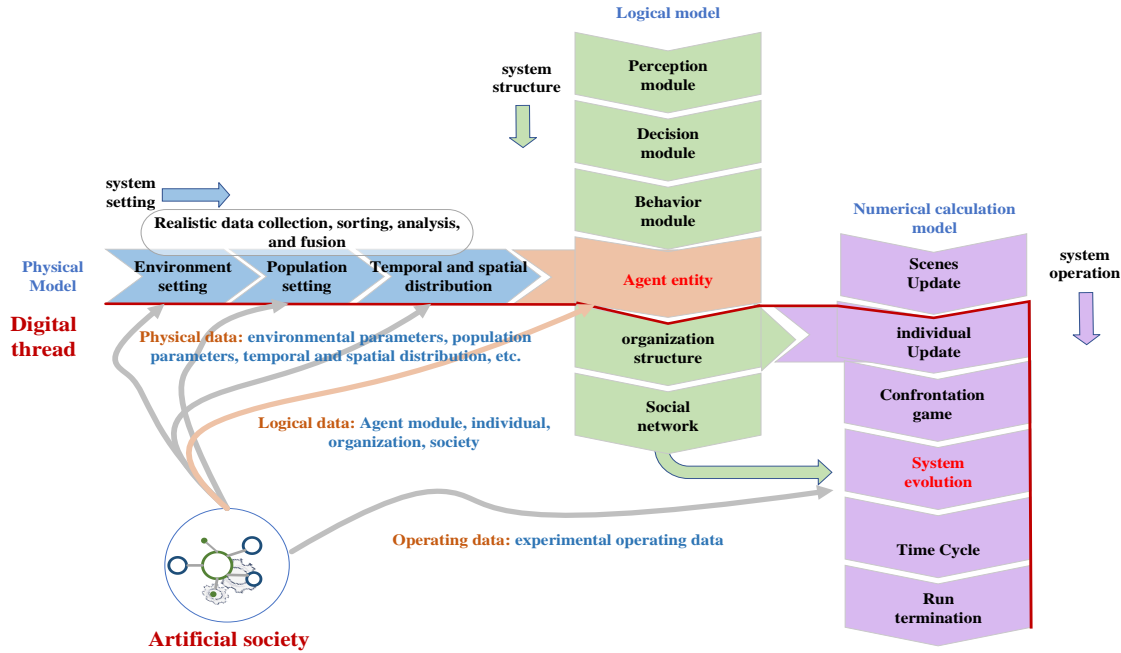

Fig. 8 Digital thread of computational experiment

*3) Interruption mechanism*

The running process of the computational experiment is the result of the interaction of internal and external factors, including the game between various individuals, and the game between individuals and the environment. External environmental factors are objective and uncontrollable, mainly including initial conditions and the external environment. Internal factors are controllable and adjustable, mainly including the organization form, cooperation strategy, coordination mechanism between individuals, and so on. Without the intervention of external factors, the experimental system will evolve naturally, which can be used to analyze the role of the initial setting and internal mechanism in the system evolution. If external intervention is applied to the experimental system, it can be used to evaluate and optimize the interventions. In order to make the experimental system evolve in the expected direction, it is necessary to implement reasonable interventions. As shown in Fig.9, a closed-loop is formed by connecting the intervention strategies and artificial society (including all kinds of agents directly and indirectly affected by the intervention). The effect of intervention strategy can be tested by adjusting the initial setting of external input and artificial society model.

According to the intervention scale, the intervention strategy can be divided into three categories: **(1)** The intervention strategies are loaded into the Agent model of artificial society (AIL, Agent in the Loop). In the process of simulation, it may affect the characteristics and behavior rules of individual agents. This mode is often applied to driverless strategies by connecting real controllers and virtual-controlled objects [86,87]. **(2)** The intervention strategies are loaded into the organization model of artificial society (OIL, Organization in the Loop). In the process of simulation, it may affect the interaction mode and learning rules of the group. **(3)** The intervention strategies are loaded into the social model of artificial society (SIL, Society in the Loop). In the process of simulation, it may affect the diffusion mode and equilibrium state of society. The artificial society will respond to the intervention and output the final results. Intervention makers compare their goals and preferences with experiment results to judge the feasibility of the intervention. After repeated trial and error, iteration, and refinement, a consensus on how to intervene can be reached.

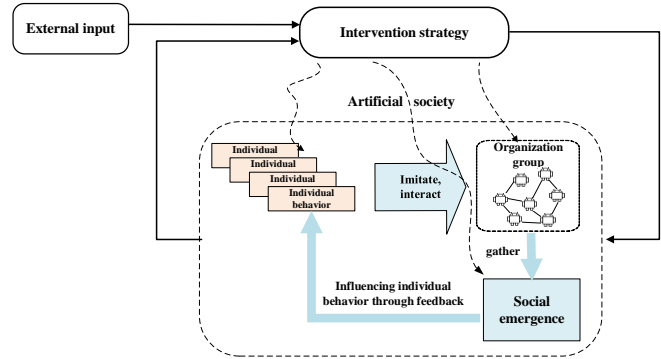

Fig 9. Interruption mechanism of computational experiment

*C. Design of experimental program*

With the increasing complexity of research objects, the number of influencing factors and their combinations are huge. Moreover, there may be associations between variables. In the computational experiment with many factors, if any combination of these factors is tested and observed, the number of experiments will increase exponentially, which cannot be available for large scale. Only through reasonable experiment design, can the ideal experimental results be obtained by the most rapid and economical method. This section describes how to design the computational experiment reasonably, including design principle, design selection, and numerical generation.

*1) Design principles*

The design of the computational experiment can be represented by the model shown in Fig.10. Experimental processes can often be visualized as a combination of operations, models, methods, people, and other resources. It transforms multiple inputs (usually a combination) into an output with one or more observable response variables. Wherein, $x_1, x_2 \ldots x_m$ are the input of artificial society system; $y_1, y_2 \ldots y_n$ are the output of artificial society system; $u_1, u_2 \ldots u_p$ are controllable factors or decisions; $v_1, v_2 \ldots v_q$ are uncontrollable factors or events. The purposes of computational experiments are: **(1)** to determine the factor collection that influences the system output through the computational experiment; **(2)** to determine the most effective controllable factor $u_i$ through computational experiment to make the output result closer to ideal level; **(3)** to determine the collection of controllable factor $u_i$ through computational experiment to make the incontrollable factor or event $v_i$ affect the system at least.

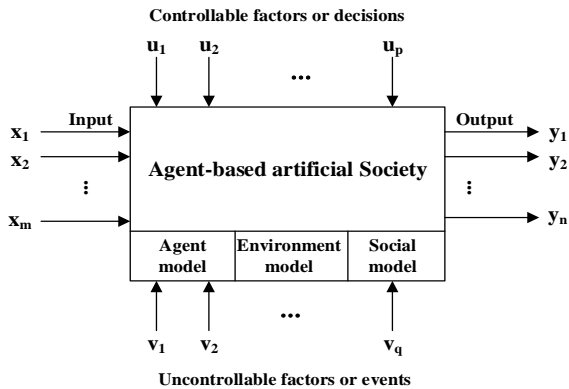

Fig.10 Design model of computational experiment

Because of the difference between physical experiments and computational experiments, many assumptions and boundary conditions in physical experiments are not satisfied in computational experiments. In physical experiments, it is usually assumed that the errors are independent and equally distributed, and even conform to the normal distribution. However, these assumptions are often not satisfied in computational experiments. Generally, the independence of errors can be guaranteed by different pseudo-random sequences. However, it is still difficult to guarantee the identical distribution of experiment errors. Therefore, many classical design methods of the physical experiment cannot be directly applied to computational experiments. For example, it is necessary to assume that no interaction effect exists between factors or only a low-order interaction effect in partial factorial design; however, for complex computational models, satisfying this assumption is difficult.

Using scientific methods to design experiments is a prerequisite for effective experiments. The statistical design of the experiment is very convenient to collect data suitable for statistical analysis and draw effective and objective conclusions. When the problem involves data affected by experimental errors, only statistical methods are objective analysis methods. As far as computational experiments are concerned, the experiment design needs to follow three basic principles of physical experiments: Randomization, Replication, and Blocking [88].

*2) Design selection*

There are many feasible design schemes for one computational experiment. To select the most appropriate solution, a cause-and-effect diagram (also called a fishbone chart) is often used as an available tool for information organization. As shown in Fig.11, the effects or response variables of interest are drawn on the fish spine, and potential causes or design factors are arranged on a string of ribs. On this basis, the selection of experimental design is carried out to define experiment purpose, the number of experiment repetitions, the appropriate order of the experiment, whether to divide the group, or whether other randomization restrictions are involved, etc.

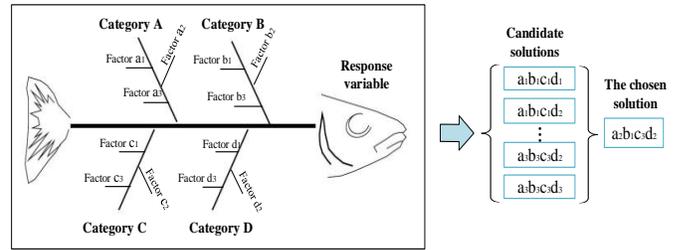

Fig. 11 Design selection model of the computational experiment

The whole process consists of the following steps: **(1)** Identification and presentation of problems: In order to use statistical methods to design and analyze experiments, researchers need to have a clear understanding of what the problem is, how to collect data, and how to analyze data. **(2)** Selection of response variables: When selecting the response variables, researchers should make sure that the variable can provide useful information on the research process. **(3)** Selection of factors: Many factors involved in the computational experiment are specified, including design factor, holding constant factor, and hate factor. **(4)** Selection of level and scope: Once the designer factors are determined, the researcher must choose the range of these factors and their specific levels, and must consider how to control them on the desired values and how to measure them.

If the interaction between different factors exists, the appropriate way to deal with multiple factors is a factorial experiment [88]. This experiment strategy is that all factors change together, not one at a time. Assuming that only two factors are considered and each factor has two levels, $2^2$ rounds of factorial experiments can help the researcher to study the individual effects of each factor and determine whether the factor has interaction. Generally, if there are k factors and each factor has two levels, $2^k$ rounds of factorial experiments should be carried out. The important characteristic of the factorial design is that experimental data can be used efficiently. In general, if there are 5 or more factors, there is usually no need to test all possible combinations of factor levels. The fractional factorial experiment is the deformation of the basic factorial design, and only one subset of all combinations needs to be tested.

*3) Numerical generation*

Simulation data will be continuously generated during the operation of the computational experiment. It becomes a key problem to be solved to decipher what kind of numerical generation strategy can make the data set as close as possible to the real situation. Synthetic data sets allow researchers to test intervention strategies on series equivalent to thousands of historical years and prevent overfitting to a particular observed data set. Theoreotically, these synthetic data sets can be generated via two approaches: resampling and Monte Carlo. Fig.12 summarizes how these approaches branch out and relate to each other[89].

Resampling consists of generating new (unobserved) data sets by sampling repeatedly on the observed data set. Resampling can be deterministic or random. Instances of deterministic resampling include jackknife (leave-one-out) and cross-validation (one-fold-out). Instances of random resampling include subsampling (random sampling without replacement) and bootstrap (random sampling with replacement). Subsampling relies on weaker assumptions, however, it is impractical when the observed data set has a limited size. Bootstrap can generate samples as large as the observed data set, by drawing individual observations or blocks of them (hence preserving the serial dependence of the observations). The effectiveness of a bootstrap depends on the independence of the random samples, a requirement inherited from the central limit theorem. To make the random draws as independent as possible, the sequential bootstrap adjusts online the probability of drawing observations similar to those already sampled.

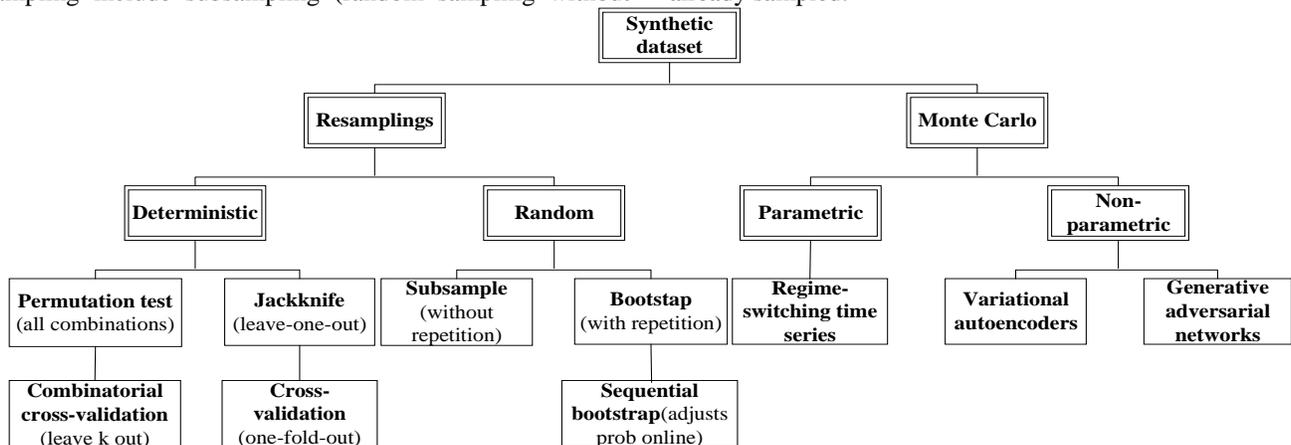

Fig. 12 Method of generating data sets for computational experiments

The second approach to generating synthetic data sets is Monte Carlo. A Monte Carlo randomly samples new (unobserved) data sets from an estimated population or data-generating process, rather than from an observed data set (like a bootstrap would do). Monte Carlo experiments can be parametric or nonparametric. An instance of a parametric Monte Carlo is a regime-switching time series model. This parametric approach allows researchers to match the statistical properties of the observed data set, which are then replicated in the unobserved data set. One caveat of parametric Monte Carlo is that the data-generating process may be more complex than a finite set of algebraic functions can replicate. When that is the case, nonparametric Monte Carlo experiments may be of help, such as variational auto-encoders, self-organizing maps, or generative adversarial networks [67]. These methods can be understood as non-parametric, non-linear estimators of latent variables (similar to a nonlinear PCA).

### D. Analysis of experimental results

In 2011, Judea Pearl found that statistical relevance cannot replace causality in the study [5]. Compared with the correlation dilemma of deep learning, the carefully designed computational experiment provides a means to discover the causality law between factors and events. The main advantage lies in the algorithmization of counterfactuals. In computational experiments, a variety of controlled trials can be set up concurrently to change the parameters of artificial societies purposefully, or to impose specific interventions. By performing a large number of repeated experiments, the experimental results are compared to explore the deep causes for complex phenomena. Generally, computational experiments can be analyzed from three levels: process analysis, scenario analysis, and what-if analysis.

#### 1) Process analysis

A main objective of computational experiments is to provide a scientific explanation of the emerging complex phenomena based on the initial or driving conditions. A typical feature of a scientific theory is that it must contain causal narratives that relate antecedents (inducement) and consequences (influence). Computational experiment theory explains the complex phenomena that emerged in the experiment system according to causality, which is characterized by the empirical model to explain the observed facts or available data. Theoretically, the event function can represent exact causal logic in detail and explain how to produce an integrated event.

However, how do event functions exist? How do different event functions explain the occurrence of integrated events? How is the probability of an integrated event according to the event function determined? To answer these and other similar questions, the logic of system complexity must be viewed at the micro-level. The sequential tree in Fig.13 provides the first-order representation of Simon's theory [90]. The main conclusion is that various results in $\Omega$ space can be produced by combination.

Event $E$ occurs in a specific environment at the initial time point $r_0$. At the subsequent time point $r_1$, individual agents can decide their adaptive change according to restricted rationality to meet the challenges of the current environment, i.e. event $D$. If they don't make a change (i.e. Event $\sim D$), the result is $E$. If they decide to adapt, they can choose whether to actually implement decisions and adaptation measures (i.e. Event $A$). If they fail to execute the event (i.e. Event $\sim A$), the result $E*$ will be produced. $E \approx E*$ can be demonstrated. If the individual has taken certain actions, the feedback may be valid or invalid at time $r_3$. If the feedback is valid (Event $W$), the result is $C$ with higher complexity. If the feedback is invalid (Event $\sim W$), environmental impacts still need to be tolerated when the result fails (Result $E**$). $S(E**) > S(E)$ can be demonstrated, and $S(X)$ represents the influence or unavailability related to Event $X$.

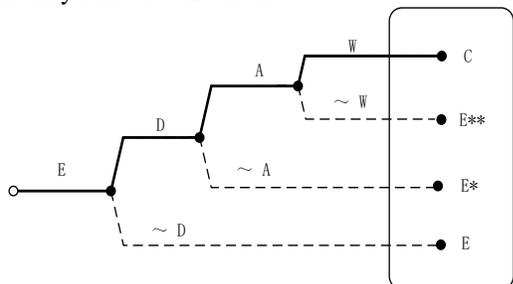

Fig. 13 Causality logic tree upon the occurrence of system complexity

Under the sequential logic model [91], the occurrence of the system complexity $C$ is explained as one branching process that experiences multiple games and resolutions in the sample space $\Omega$, i.e. one of the possible events. As shown in Fig.13, the emergence of system complexity $C$ requires at least four necessary sequence conditions, which have a significantly low probability. The occurrence probability of other results (fail $E$, $E*$, and $E**$) is relatively high. In conditional logic mode, when the system complexity occurs as an integrated event, the background can be studied according to the current site and the necessary or sufficient conditions can be explained. Generally, the system complexity $C$ occurs in a dual causation mode: depending on the simultaneous occurrence of necessary conditions (events $X_1, X_2, X_3, \ldots X_N$ are connected via AND); or, depending on the occurrence of any sufficient condition (event $Z_1, Z_2, Z_3, \ldots Z_M$ are connected via OR).

*2) Scenario analysis*

The basic idea of scenario analysis is to acknowledge that the future outcome of the event and the way to achieve this outcome are both uncertain. It focuses on the prediction of possible scenarios so that researchers can effectively respond to possible scenarios [92]. In essence, it is to complete the description of all possible future trends, including three parts: **(1)** Describe the process of event occurrence and development, and analyze the dynamic behavior of event development; **(2)** In the complex "event group", sort out key elements and event chain through induction and organization; **(3)** Obtain the dynamics and relevance of the time scenarios embodied in this process, and establish a logic structure for the same events. The traditional means for scenario analysis depends on artificial imagination and reasoning. Computational experiments introduce quantitative methods for scenario analysis and construct the "scenarios" of occurrence, development, transformation, and evolution of events. Under the action of situational factor, the probability of transition between various scenarios is affected, and the selective correlation transitions may occur between various scenarios. From the development and evolution of the scenario, it can be described as follows:

$$S = <T_r, Des, Actor, In, Action, Out, ES> \quad (2)$$

Wherein, $T_r$ indicates the trigger condition for scenario initiation; *Des* is a brief linguistic description of the scenario; *Actor* is the participants involved in the scenario; *In* is the data or condition obtained from the external environment, which changes with the external environment; *Action* is the behavior or action executed after the scenario starts; *Out* is the data generated in the execution process of scenario behaviors; *ES* is the alternative scenario when the abnormal situation occurs. The scenario options constructed with formula 2 can cover the various uncertainties of the system, thereby establishing the relationship between the evolution of artificial society and the actions of multiple agents.

As shown in Fig.14, the traditional metrological regression methods can only describe the statistical relationship from scenario 1 to scenario 2 (arrow 1), while "agent analysis technology" explains the relations between pre-intervention scenario 1 with post-intervention scenario 2 by analyzing micro-individual behaviors. Such analysis is mainly based on the behavior analysis of intelligent algorithms: **(1)** the agent behavior pattern is explored from massive agent behavior data (Arrow 2); **(2)** the agent behaviors are self-learned (Induction-probability model) and self-evolved (Deductive-Rule Model) based on intelligent algorithms (such as neural network, genetic algorithm, tabu search, simulated annealing, etc.); **(3)** the evolution behavior rule of the agent is used to explain or predict the future phenomenon 2 (arrow 3).

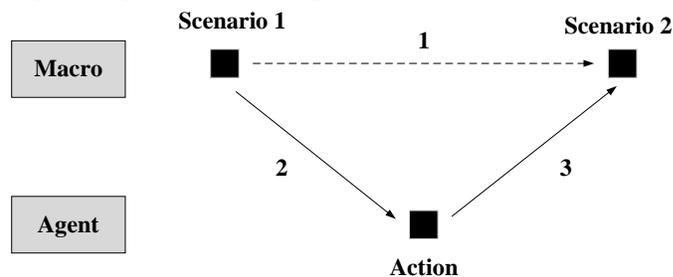

Fig. 14 Scenario analysis model based on agent behavior

The behavioral analysis methodology assumes that the collective behavior of agents is used as an intermediary between the source phenomenon 1 and the prediction phenomenon 2. In other words, the behavior chain between the two phenomena can be replaced by intelligent algorithms. However, it can only solve the descriptive problem of the macro phenomenon, but not explain the emergence of complex social phenomena. The agent behavior based on intelligent algorithms is based on statistics. The law of behavior produced by this method may not exist in reality; and, even if it exists, it cannot express the "causality" from phenomenon 1 to phenomenon 2.

*3) What-If analysis*

When using scenario analysis, if the micro-behavior assumption of the agent and the formal features produced by the model are both similar to the real world, we generally think that this model can reasonably represent the laws of the real world. There are several problems with this logic: "a mechanism can produce a phenomenon" is far from the conclusion that "the mechanism is the cause of this phenomenon". Different Agent models can produce the same formal features. So, which is the real internal mechanism that produces these features? In order to find the causality between variables, the what-if analysis generally adopts the three conditions proposed by John Stuart Mill (British philosopher): **(1)** Association: Two events must be co-varying or changing together; **(2)** Direction of Causation: An event must occur before another event; **(3)** Confounding variable. If this computational experiment (usually a series of experiments and research processes) proves that the three conditions are satisfied at the same time, then it can be said that the causality is credible and valid under certain conditions (experimental conditions).

In the computational experiment, what-if analysis mainly depends on mechanism-based behavior analysis technology. "Mechanisms" refer to the generation rules of agent behavior and the interaction rules between agents, as well as the internal mechanism of specific macro results. As shown in Fig.15, mechanism-based behavior analysis is the process of analyzing the causality chain. The emphasis of the analysis is the agent, especially the orientation to the action of the agent: **(1)** Analyze the influence of hypothesis 1 on the orientation to action, including beliefs, desires, and intention (BDI), etc. [93], i.e. the analysis of the causal chain between hypothesis 1 and the orientation to action (arrow 2); **(2)** Analyze the mechanism between the orientation to the action of agents and other environmental factors, i.e. the generative model of agent behaviors (arrow 3); **(3)** These rules will be manifested as the behavior of the agent in the agent-based model, and then the deduction results of hypothesis phenomenon will emerge (arrow 4).

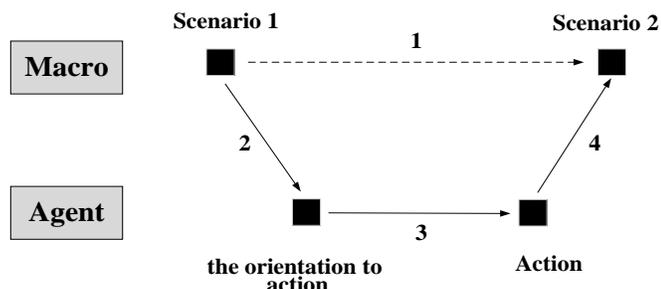

Fig. 15 What-if analysis model based on mechanism

The mechanism-based behavior analysis is an analysis of the "influence-action-emergence" mechanism from macro phenomena to micro behaviors to macro phenomena, rather than a description of the coverage law between macro phenomena based on statistics or intelligent algorithms. As a result, it can penetrate every aspect of the evolution of artificial society and restore the mechanism of intervention. Therefore, the what-if analysis is the core of the computational experiment. To further quantify the causal relationship between the two nodes, the Probabilistic Causal Model (PCM) [94] can be used to formalize and define relevant operations. By adopting a probabilistic causality model and do-calculation axiom system, the distribution of another variable can be concluded when one variable is intervened. In this way, the causal relationship between two variables can be quantified.

### E. Verification of experimental results

Experimental verification is used to examine whether the model accurately maps the prototype system in its application domain, to ensure that the model meets the application purpose. To enhance the confidence of the model, credible proof must be provided to ensure that the model hypothesis represents the laws of a real system. High confidence is the basis on which computational experiments work. Only when breakthroughs are made in the model verification can the computational experiment method become a powerful tool to understand the operation laws of complex systems. At present, various methods can be used to verify the validity of the computational model [95]. As shown in Fig.16, the experiment verification can be divided into three parts according to the experiment operation process: structural verification, data verification, and result verification.

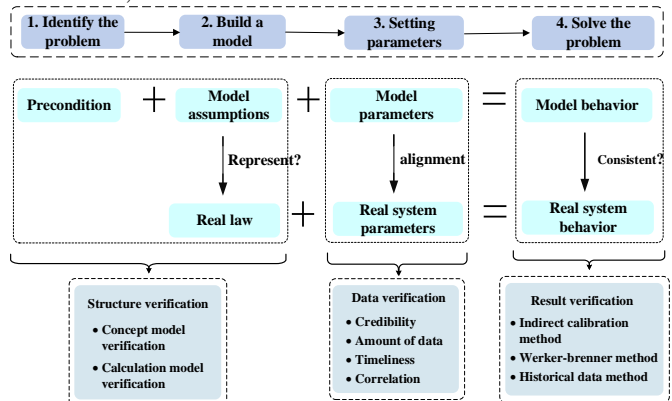

Fig. 16 Classification of experiment verification

### 1) Structural verification

The conceptual model is an abstraction of the real system and can be implemented through programming to obtain a practically executable computational model. Structural verification is to verify the process of model construction so that the model can reflect the internal structural characteristics of the real system, and ensure that the intermediate process of model behavior is correct. In other words, the hypothetical model is a sufficient condition to produce the behavior result. The modeling process of artificial society includes conceptual model design and computational model implementation. Therefore, the structural verification also includes conceptual model verification and computational model verification:

### 1.1) Conceptual model verification

In this stage, the consistency of conceptual models with the research purposes, given assumptions, existing theories, and evidence is examined to determine that the simplified processing of the model will not seriously affect the credibility of the model and the understanding of the important characteristics of the real system. At the same time, the conceptual model should satisfy inherent integrity, consistency, and correctness. Whether the model is reasonable can be judged three types of methods: expert judgment, theoretical comparison, and empirical data fitting, which are often qualitative and not strictly proven. By comparing with other existing models, researchers can greatly improve the maturity of conceptual models in the following aspects: **(1)** Describe the application scope and defects of the model; **(2)** The mainstream theory uses the same assumptions, or other researchers made similar studies and assumptions; **(3)** Develop and analyze a set of models with the same core assumptions but different additional assumptions; **(4)** Establish an equation model with similar characteristics to the behavior of the original Agent model.

### 1.2) Computational model verification

In this stage, it is mainly to check whether the algorithm, coding, operating conditions, etc., are consistent with the conceptual model. The conceptual model needs to be transformed into a computational model that can be run by a specific computer system through programming. This process involves a variety of factors and undetectable errors. Because the Agent model is discrete, and significant differences exist between various programming languages and computer systems, the programming details have an important impact on the model results. At present, the main methods used to verify the computational model are: (**1**) Check the consistency between the computational model and the conceptual model; (**2**) Run the same code on different computers, different operating systems, and different pseudo-random number generators; (**3**) Use different programming languages to implement and compare conceptual models; (**4**) Test the behavior of the computational model in extreme cases. Code checks are involved in any case. In principle, code checking can detect all programming errors.

*2) Data verification*

In computational experiments, the model behavior obtained by model assumptions and model parameters is usually regarded as a prior condition for experiment research. The building of the model is actually equivalent to proposing a hypothesis, and the builder believes that this hypothesis represents the law of the real system. By performing computational experiments and observing the model behavior results, a causal explanation of the real system is made. The quality of the input data will directly affect the credibility of the final experimental results. Poor data input can even lead to great deviations from the actual situation. It is necessary to guarantee the data availability from the source to build an effective model. Many researchers [96,97] have noted the problem of data quality, including consistency, precision, completeness, timeliness, and entity identity. For computational experiments, the following factors need to be considered:

**Credibility:** Data collection is actually an artificial process of collecting data according to propositions. Therefore, it has a tendency at the beginning, and the operation process cannot be guaranteed to be completely true. Since there is no absolute "truth", data that strictly follows scientific principles and is less subjective is closer to "true" and credible. In general, government and authority data are more reliable.

**Data size:** In statistics, the sample size is directly related to the accuracy of the inferred estimates. In the case of a given population, the larger the sample size, the smaller the estimation error. On the other hand, the smaller the sample size, the greater the estimation error. In estimating the parameters of the computational model, for the same population (as in a region), larger data size can reduce the probability of random errors and the problem of underrepresentation.

**Timeliness:** The operation of a complex system is a dynamic process. Data trends may fluctuate sharply as scenarios and external interventions change. Therefore, the timeliness of data is an important factor affecting the experimental results, especially for some projects with strong timeliness. There is an inevitable problem in that it takes time to collect and acquire data. In particular, in the early stages of some emergencies, the problem of data delays is very prominent. For the computational model, the better the timeliness of the data, the better the experimental effect.

**Relevance:** In many cases, there is no data we want to study in the data resources. In other words, there is a lack of direct data related to the research problems. Hence, the researchers will have to find alternative variables for research. The correlation between alternative variables and direct variables will affect the experimental results.

*3) Result verification*

The result verification is mainly used to evaluate whether the experimental results are consistent with real system behavior. Real data can be collected from the real world, but the data output from the model is generated by the set model operating mechanism. The data generated by the model is compared with the real data to infer whether it "appropriately" reflects reality, to achieve the purpose of result verification. In recent years, the mainstream result verification methods include the indirect calibration method [98], the Werker-Brenner method [99], and the historical data method [100]. However, the artificial society model is only a finite sample of the population after the finite operation. The model behavior is unstable and parameter sensitive. Therefore, compared with the real system, the result verification of the artificial society model will face the following problems:

*Q1: How to deal with parameter combination explosion and sensitivity analysis?*

A number of model parameters will lead to combination explosion. Even if the experimental results are consistent with the real data, it is difficult to determine the decisive factors affecting the experimental data. At the same time, the model results are not stable and are largely influenced by the initial conditions. At present, researchers basically adopt two methods to reduce the parameter space. The first method: at the beginning of modeling, the parameter space is reduced by existing data, facts, corresponding models and theories, expert experience, and knowledge. This method can make full use of existing knowledge, but it is easy to introduce unverified prior information. The second method is to perform sensitivity analysis on formatted features after the model is established and run, to find and further analyze sensitive parameters, thereby reducing the parameter space. This idea is very convincing when using experimental design, statistical method, and Monte Carlo method. But it is difficult to implement them in many cases. These two methods are often combined to reduce space in two directions.

*Q2: How to compare model output and empirical data?*

The result of running the model is the probability distribution of multiple samples, while real-world data is only one sample. It is difficult to test a probability distribution with a sample. The model output can be considered "consistent" with the real data to some extent. At present, it is mainly confirmed subjectively. The results are largely restricted by the subjective consciousness of the analysts. This judgment must consider not only the consistency of model output but also the consistency of input and output. That is, the model needs to be calibrated before the consistency of the results is considered.

*Q3: How to validate results that have not occurred in reality?*

Under normal circumstances, the computational model is verified using the existing empirical data. Sometimes, the results of computational experiments may not have been produced in the real world. In this case, we need to make full use of domain knowledge to find out what causes this phenomenon and whether it is a trend in the real world. Additionally, we need to decide whether the model is effective. The results are largely restricted by the range of knowledge and problem analysis capabilities of the analysts.

## IV. APPLICATION CASES

Computational experiment, as a new scientific research methodology, can be applied in three aspects: thought experiment, mechanism exploration, and parallel optimization. This section will give some classic examples to explain the concept of the computational experiment method.

### A. Thought experiment

The thought experiment does not model a specific scenario or specific real social system but pursues the abstract logical relationships that describe general social systems. It is hoped to explore and quantitatively analyze the unpredictable results of certain hypotheses on human society through experiments. This kind of research avoids the mapping problem from real society to artificial society. This type of research can often give metaphors, enlightenment, and qualitative trends, rather than precise answers to complex questions.

*1) Sugarscape model*

In 1996, Joshua Epstein and Robert Axtell came up with an "artificial society" model - Sugarscape, which can be used to carry out relevant experiments in economics and other social sciences [45]. As shown in Fig.17, the SugarScape model is a closed world composed of grids: the red dots indicate Agents that can only walk in this world; the yellow part indicates social wealth - sugar and yellow concentration indicates the sugar distribution concentration. Each Agent contains three properties: range of vision $r$, metabolism of resources $v$, and sugar quantity $s$. Agents walk according to the following rules: **(1)** All cells in the range of vision $r$ are observed and the cells with the largest sugar content are determined as the target; **(2)** If there is more than one cell with the maximum sugar content, choose the nearest cell; **(3)** Move to the grid; **(4)** Collect the sugar of the cell and update the corresponding variables $s$.

As shown in Fig.17, as the experiment progresses, some agents may obtain more sugar because of their unique personal ability, the resource advantages of the location, and so on; sugar-deficient individuals will die so that most Agents will be gathered in two regions with higher sugar concentration. Ultimately, a few Agents have a large amount of sugar, while most Agents have only a small amount of sugar, which verifies the famous Matthew effect in social science. Further, by adding a variety of resources (such as spices) to Sugarcape artificial society, how individuals form markets through resource exchange in real society can be studied. Furthermore, we can study social phenomena such as environmental change, genetic inheritance, trade exchanges, and the market mechanism by changing the different rules that Agent follows.

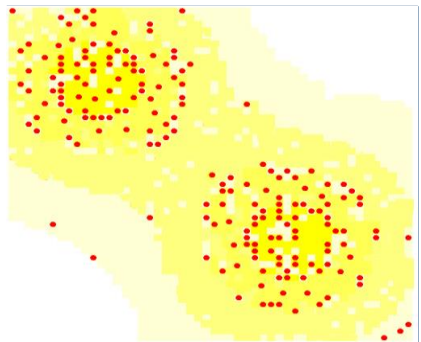

Fig. 17 Distribution of sugar and Agent in Sugarscape

*2) Schelling model*

The Schelling model (Schelling isolation model) was proposed by Thomas Schelling (an American economist). The model describes the influence and function of homogeneity on spatial segregation and reveals the principles behind racial and income segregation [60]. The model contains three elements: Agent that generates behavior, behavior rules to be followed by Agent, and macro results caused by Agent behaviors. As shown in Fig.18, the whole city is seen as a giant chessboard in the experiment. Each small grid on the board allows Agent to live or idle. There are two kinds of Agents (red and blue) of equivalent number and about 10% of grids are blank (green). Each Agent has the lowest threshold. Once the number of neighbors of the same kind falls below the threshold, they migrate to an unoccupied location that meets their residence requirements. The behavior rules of Agents include: **(1)** Calculate the number of its neighbors; **(2)** If the number of neighbors is greater than or equal to their preference value, Agent thinks it is satisfied and stops moving; otherwise, continues moving; **(3)** Agent will find a blank grid that satisfies its preference value and is closest to it and move there.

As shown in Fig.18, as time passes, the degree of isolation between different kinds of agents will eventually show a very obvious state. Through changing experimental settings (such as Agent lifetime value, blank spaces, etc.), it can be found that changing tolerance is not enough to avoid apartheid, because almost every agent pursues the same kind of neighbors. This phenomenon will trigger our thinking about social issues: To what extent will racism transform the whole society into such an isolation mode? How can apartheid be reversed? The researchers extended the model and applied it to the study of ethnic conflicts in different regions and obtained better insights [101,102].

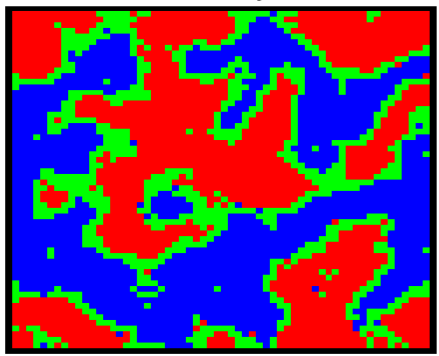

Fig. 18 Apartheid experiment based on the Schelling model

*3) Landscape theory*

Robert Axelrod (University of Michigan) proposed the landscape theory to study the alliance in human society and predict the equilibrium [103]. The theory treats social units as particles in the physical world. They may be drawn together by some attraction and separated from each other by exclusion. This pull-and-push environment brings particles together in different combinations. In various combinations, the sum of the attraction and exclusion effects of each group member is called "total energy". The energy landscape is a graph that represents all possible combinations and corresponding energies. The more stable the combination, the lower the energy. Axelrod and his colleagues simulated the possible coalition camp on the eve of the Second World War by using the landscape model. The interaction between two countries is judged by using six factors, including ethnic situation, religious beliefs, territorial disputes, ideology, economic situation, and history. A simple weighting factor is set for each factor. The weighting factor is +1 when converging, and is -1 when conflicting and antagonizing.

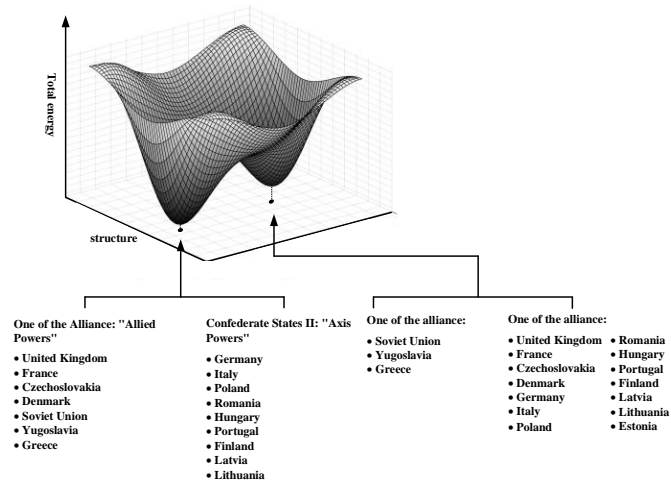

Fig. 19 Landscape map of coalition camp on the eve of the Second World War

Fig.19 is the landscape map of all combinations. There are two basin structures on the map (namely, the minimum energy value) : a deeper basin and a slightly shallower basin. : The deeper sinking basin is close to the actual alliance of allies and axis powers. Only Portugal and Poland are classified as the "wrong camp". Another basin predicts a very different scenario: the Soviet Union confronted all other European countries. According to the energy landscape, which "valley bottom" Europe will eventually slide to is determined by the location of the starting point. The landscape theory divulges that for the special case of the alliance, it is possible to carry out practical calculations and obtain credible values and similar results. The study of history becomes more solid through the computational experiment of the landscape model. In this way, we have a certain degree of quantitative language to discuss the world situation. This model can help us identify the influencing factors in the historical development process, make the scope of research more clear, and achieve the purpose of prediction.

*B. Mechanism exploration*

The mechanism exploration is a modeling of a real social system, with emphasis on high matching between artificial society and real social systems. It is expected to solve problems that exist or may exist in the existing society. Such applications often encounter the validity verification problem of mapping from real society to artificial society. At present, researchers hope to solve such problems through big data acquisition and processing technology, so that the experimental results can solve real problems.

*1) Artificial stock market*

The stock market is clearly a complex system, which shows complexity and unpredictability with multiple elements and hierarchy. To demonstrate and understand how investors make portfolio choices, Santa Fe Institute (SFI) proposed the "Artificial Stock Market" model (ASM) in 1987, which understood the complexity of the economic system from a new perspective [46]. In this model, the hypothesis of a completely rational and all-knowing "economic man" is discarded, and replaced by a bounded rational man who can learn and adapt to the environment and make decisions using inductive methods. The economic system is seen as a complex system of interacting individuals. In this visual market, several trading Agents make predictions by observing the changing stock prices in the digital world and make decisions on whether to buy stocks and the purchased quantity to maximize their utility. All agents develop their expectations independently and have the ability to learn. Their decision adjustment can be made according to the success or failure in predictions. In turn, the decision of all traders determines the competitive state of demand and supply and then determines the price of stocks.

As shown in Fig.20, the whole stock market constitutes a self-closed computing system. The running time of the model is discrete and the experiment can be carried on indefinitely. ASM provides a good metaphor for the real stock market, which can be used to prove an effective market theory, i.e. different expectations eventually evolve into the same rational expectations. More complex behavior that occurs in the agent and organization levels can support the views of real investors and explain the real phenomenon of financial markets. Subsequently, several researchers have improved the ASM model to conduct deeper analysis, such as linking micro-level investor behavior with macro-level stock market dynamics, the effectiveness of price limits, and so on [104,105].

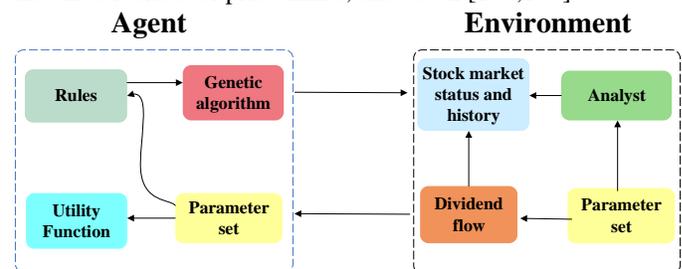

Fig. 20 Agent and stock market interaction structure

*2) Political mechanisms*

Claudio Cioffi-Revilla (George Mason University) proposed the MASON RebeLand model to analyze how the state of a polity or its political stability is affected by internal (endogenous) or environmental (exogenous) processes, such as changing conditions in its economy, demography, culture, natural environment (ecosystem), climate, or combined socio-natural pressures [17]. Figure 21 shows a "map" view of RebeLand as a polity or country that is situated in a natural environment. The country itself consists of an island surrounded by water, therefore omitting external or neighboring interactions with other countries. The RebeLand environment consists of terrain and a simple weather system that simulates climate dynamics (e.g. prolonged droughts, climate variability, and so on). The RebeLand political component comprises a society and a system of government for dealing with public issues through public policies. Initially, the government formulates policies to address issues that affect society. Later in the simulation, under some conditions, the society can also generate insurgents that interact with government forces, as well as other emergent phenomena.

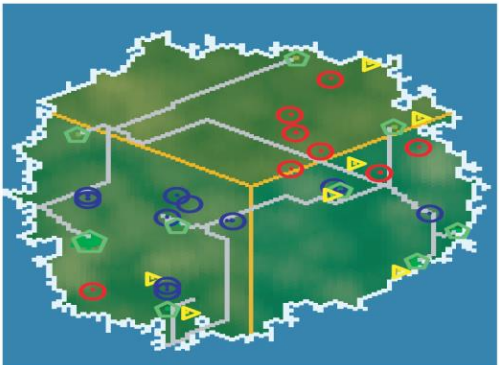

Fig.21. Map of RebeLand Island showing its main natural and social features. Cities are shown in green, natural resources in yellow, and rebel and government forces in red and blue, respectively. Roads and provincial boundaries are in gray and yellow, respectively. Physical topography is shown on a green-tone scale and the island is surrounded by ocean.

This study demonstrates three political states: stability, instability, and failure. An important result pertains to the general overall resiliency of a polity, normally requiring not just one or a few stressful issues to experience polity failure, but a large set in combination (such as inflation plus insurgency plus environmental stress). Such results may not immediately yield actionable policy recommendations in terms of specific programs, but at the very least they may offer new insights of value to both researchers and policy analysts. Such studies have also been used in electoral forecasting and applied political science [106].

### 3) Epidemic spread

With the prevalence of sudden infectious diseases (e.g. SARS, H1N1, COVID-19, etc.), the transmission model of infectious diseases has become a hot topic in research field. However, the spread of infectious diseases is a dynamic and uncertain process. There are not only multiple influencing factors but also many unknown fields and uncertainty, putting much pressure on predictive research. The artificial society model is used to realize the integration of basic data, model method, and analysis results in the computational experiment, which involves three aspects: the generation of the epidemic situation (such as the number of initial infections, exposure rate, transmission rate, virus incubation period, mortality rate, recovery probability, etc.), the representation of spatial geographical features (such as urban type, transportation network, population density, temperature, weather, urban infrastructure, etc.), and the modeling of resources and governance capacity (such as medical resources, social organization, prevention and control measures, information transparency, etc.). Table 3 compares the characteristics of several artificial societal systems related to infectious diseases. Different artificial society systems have their characteristics in implementation, representation, and accuracy.

Table 3 Comparison of the characteristics of several artificial society systems related to infectious diseases

| Features | BioWar[15] | EpiSimS[76] | GSAM[107] | CovidSim[108] | ASSOCC[109] | SlsaR[110] |
|---|---|---|---|---|---|---|
| Disease type | Droplet spread, Physical contact spread | Smallpox, influenza | No specific disease (taking H1N1 as an example) | COVID-19 Other respiratory viruses | COVID-19 | COVID-19 |
| Main application | Effect evaluation, Strategy optimization | Effect evaluation, Strategy optimization | Study on the spread and control of infectious diseases | Effect evaluation, Strategy optimization | Effect evaluation, Trade-off policy | Assessing the costs and benefits of different intervention policies |
| Simulated scale | Medium cities in USA | Medium cities in USA | Globe-scale | Country-scale | Country-scale | Country-scale |
| Stimulation method | Multi-Agent | Multi-Agent | Multi-Agent | Geographical spatial unit | Multi-Agent | Multi-Agent |
| Development Language | C++ | - | Java | C++ | R language and Netlogo | NetLogo |
| Visualization | No | Yes | Yes | Yes | Yes | Yes |
| Open source | No | No | No | Yes | Yes | Yes (available online) |

At present, computational experiments have become an important means to study the mass spread of infectious diseases, which are mainly applied in 3 aspects: **(1)** Prejudgment of transmission trends. The developed transportation systems make it easier for major infectious diseases to spread on a large scale. When the epidemic situation has not broken out in certain areas, the analysis and judgment of its transmission trend are an important prerequisite for emergency preparedness. **(2)** Pre-assessment of impact. The outbreak and evolution of the epidemic situation in one region will directly do harm to human health condition, and will also have side-effects on social and economic environment. The quantitative assessment of its impact is an important basis for emergency reserve and intervention intensity decisions. **(3)** Intervention strategy optimization. There are a variety of interventions for emergency prevention and control of major infectious diseases. Each intervention has a different target and intensity. In practice, it is an important and difficult problem to make emergency decisions, that is, how to choose reasonable intervention measures and form a combined intervention strategy that can control the epidemic situation and reduce intervention cost.

### C. Parallel optimization

The purpose of scientific research is to provide a causal hypothesis that can link existing facts together. If the artificial society model represents this law, the output behavior of the model can be effective in the real world. Parallel optimization, through the establishment of an artificial social model that has a homomorphic relationship with the real society, realizes the parallel execution and cyclic feedback of the two, thereby supporting the management and control of the real complex system.

*1) Island economy*

Economic inequality is growing globally and is gaining attention because of its negative impact on economic opportunities, health, and social welfare. For governments, tax policies can be used to improve social outcomes. However, because of the coupling between tax and labor, taxes may reduce productivity. Therefore, how to reduce economic inequality while ensuring productivity remains a problem to be solved. The lack of appropriate economic data and experiment opportunities makes it difficult to study such economic problems in practice. For this purpose, Salesforce proposed a new study named "The AI Economist". Economic simulation through AI agents can explore tax strategies that can efficiently balance economic equality and productivity [111]. A two-level reinforcement learning framework is formed between Agent and social planner with learning and adaption functions. Fig.21 gives the core implementation framework of "The AI Economist", which emphasizes the game evolution mechanisms between Agent and Planner, i.e. how to use reinforcement learning to realize the joint optimization of agent behavior and taxation strategy.

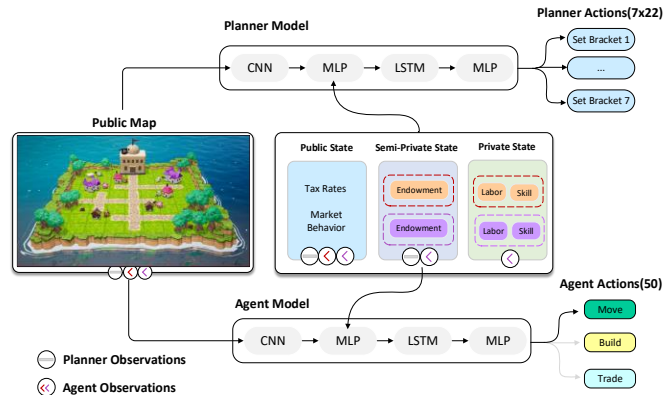

Fig. 22 The core implementation framework of "The AI Economist"

Researchers have compared the operation results of the AI Economist and the free market (Non-taxation or redistribution), United States federal tax plan, and tax strategies generated from the Saez framework [112]. The experiment shows that the AI Economist can increase the trade-off between economic equality and productivity by 16% compared with the tax framework proposed by Emmanuel Saez. The framework can be optimized directly for any socio-economic goal, which does not use any prior knowledge or modeling assumptions, and only learners from observable data. Salesforce developers hope that the AI Economist can solve the complex problems that traditional economic research cannot easily handle, and conduct objective research on the impact of policies on the real economy.

*2) Virtual Taobao*

Commodity search is the core trade on Taobao, which is one of the largest retail platforms. The business goal of Taobao is to increase sales by optimizing the strategy of displaying the page view (PV) to the customer. As the feedback signal from a customer depends on a sequence of PVs, it is reasonable to consider it as a multi-step decision problem rather than a one-step supervised learning problem. The engine and customers are the environments of each other. RL solutions are good at learning sequential decisions and maximizing long-term rewards. One major barrier to directly applying RL in these scenarios is that current RL algorithms commonly require a large number of interactions with the environment, which take high physical costs, such as real money, time from days to months, bad user experience, and even lives in medical tasks.

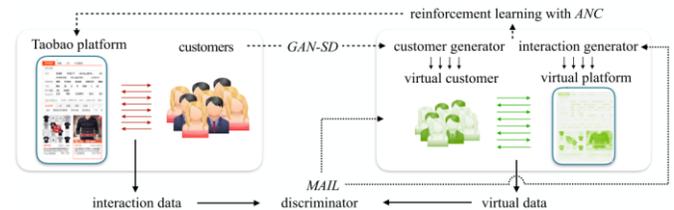

Fig. 22 Virtual Taobao Architecture with reinforcement learning

To avoid physical costs, the researchers employed simulators (i.e, Virtual Taobao) for RL training. Thus, the policy can be trained offline in the simulator [113]. In this work, Virtual Taobao is designed by generating customers and generating interactions. As shown in Fig.22, the GAN-for-Simulating-Distribution (GAN-SD) approach is proposed to simulate diverse customers including their request; the Multi-agent Adversarial Imitation Learning (MAIL) approach is proposed to generate interactions. After generating customers and interactions, Virtual Taobao is built. The experiment results disclose that Virtual Taobao successfully reconstructs properties that closely parallel the real environment. Virtual Taobao can be employed to train platform policy for maximizing revenue. And, compared with the traditional supervised learning approach, the training strategy in Virtual Taobao achieved more than a 2% improvement of revenue in the real environment.

*3) Unmanned driving*

In unmanned driving, the ability of unmanned vehicles to understand complex traffic scenarios and autonomously make driving decisions needs to be repeatedly tested and verified. It is one of the major challenges in the AI field. However, it is challenging to build real test scenarios (such as blizzard, rainstorm, typhoon, etc.) due to high costs in creation, duplication, and iteration. Therefore, the training and testing of unmanned driving strategies in the virtual world have become a feasible technical choice. It is not only controllable and repeatable but also safe and effective. In order to achieve an autonomous driving test that is infinitely close to the real world in the virtual world, three levels of reduction are required: **(1)** Geometric reduction: 3D scene simulation and sensor simulation are required to craft the environment and test vehicle conditions that mirror the real world; **(2)** Logic reduction: the decision making and planning process of test vehicles should be simulated in the virtual world; **(3)** Physical reduction: the vehicle control and body dynamics results should be simulated. At the same time, the simulation platform should satisfy the characteristics of high concurrency and realize the combination of vehicle responses in all scenarios. At present, parallel learning methods are increasingly used in virtual scene generation and intelligent testing of unmanned driving [10,11,87,88].

As shown in Fig.24, Tencent built a virtual city system parallel to the real physical world, based on a simulation platform, a high-precision map platform, and a data cloud platform. The virtual city contains not only static environmental information but also dynamic information (such as traffic and passenger flow). It can not only support the development and safety verification of self-driving, but also contribute to the construction of intelligent transportation and smart cities. In order to improve the utilization of road data and further enrich test scenarios, a large amount of road data can be used to train the traffic flow Agent. As a result, traffic scenes with high reality and strong interaction can be generated for closed-loop simulation, testing efficiency improvement, and collection cost reduction. For example, if the tested self-driving car wants to overtake, Agent AI may be used to control NPC (Non-player character) vehicles to make avoidance or other game actions consistent with the real world.

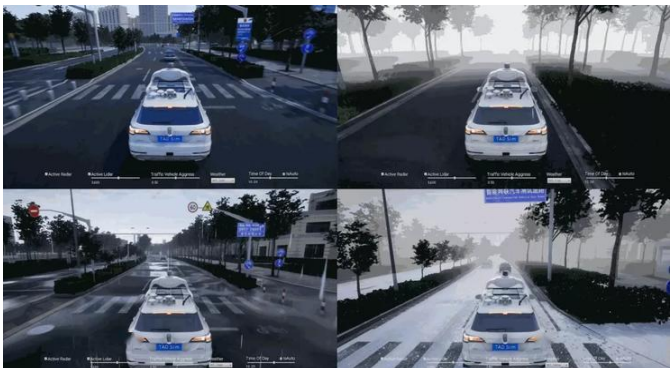

Fig. 24 Scene demonstration of Tencent unmanned driving testing system

## V. PROBLEMS & CHALLENGES

Computational experiments would cultivate human ability of observing and understanding the world and learn to deep thinking. However, compared with its increasingly widespread application, computational experiments have not been advanced as a methodology and technology. Therefore, the computational experiment method should solve the following three questions to transition into the toolbox of mainstream researchers: **1)** How to define the artificial society using big data, i.e. describing intelligence; **2)** How to predict the future using computational experiments, i.e. predictive intelligence; **3)** How to realize the feedback intervention to the real world, i.e. guiding intelligence.

### A. Q1: How to implement describing intelligence?

By abstracting and simplifying the complex behaviors and phenomena in real social systems, we can obtain an artificial society model which can run on a computer platform and satisfy logical rationality and correctness. This is the basis of computational experiments. Currently, there are two main methods of artificial society modeling: **1)** Data-based modeling method. The complex system is regarded as a black box, which focuses on the relationship between input and output, and does not model or simulate complex processes within the system. In practice, complex systems are often replaced by statistical models based on data and intelligent algorithms, including Graph Neural Network (RNN), vector machine, fuzzy method, etc. **2)** Mechanism-based modeling method. The complex system is understood and analyzed from an overall perspective. Following the principle of "simple consistency", the structure and function of each part of the system are designed and restored. In the process of building a computational model according to the bottom-up principle, the accuracy and complexity of the underlying microscopic model play an important role in the evolution of the entire complex system.

However, the complex system is characterized by diverse elements, frequent changes, coupled relationships, and multi-level/multi-stage operations. Restricted by cognitive ability, the traditional mechanism modeling is difficult to use to describe the operation and evolution mechanism of complex systems. Due to the lack of internal structure and mechanism information for process units, data-driven modeling relies heavily on the quantity and quality of data samples, making it difficult to analyze and explain the system mechanism in depth. Therefore, how to construct an artificial society model for CPSS has become one of the promising research problems in this field. Mechanism analysis is helpful to identify the essential characteristics of the system and set up an effective model structure. And, data-driven methods can automatically acquire information and knowledge hidden in data. By coupling the rule model describing individual behavior with the statistical model, a hybrid model combining system mechanism analysis and big data analysis is established. It not only reduces the dependence on data but also makes up for the situation where traditional models are unlikely to simulate individual behavior rules. In the future, there will be ample space to build virtual artificial society models with more realistic behavior.

It should be noted that due to the large freedom of the Agent model and the lack of a standardized core model, it is easy to over-promote and exaggerate experiment results, thus lead to a lot of criticism. Therefore, in the process of progressive modeling of complex systems, we should note: **1)** It is necessary to identify a simple initial model with all details that can represent the core elements of the reference system. **2)** The design of the model is not in a random sequence that must provide incremental help for the final model. **3)** Throughout the development process, model validation is necessary, but it needs to be moderate. Because the intermediate stage model is not yet mature, it may be rejected due to a lack of sufficient empirical support. **4)** It is necessary to determine the final model of the reference system, otherwise the improvement of the model will continue indefinitely.

### B. Q2: How to implement predictive intelligence?

In the complex system, each link is full of high uncertainty triggered by individual behavior, environment, and the implementation of the intervention strategy. As a result, the analysis, formulation, and implementation of intervention strategies are accompanied by the complexities of uncertainty. Big data technology is a summary of facts based on the analysis of historical data. It is helpless for uncertain events that never happened. Therefore, how to develop scientific and universally accepted intervention strategies for complex systems is a basic problem to be addressed using computational experiments.

With the help of artificial society, real-world issues can be accurately described, analyzed and controlled in the virtual world, resulting in massive system simulation operating data. Based on this, we can use computational experiments to deduce various scenarios that have never occurred in reality, discover problems and defects in system operation, find out the reasons, and put forward practical solutions or suggestions. Various forms of analysis can be adopted for the computational experiment, including process analysis, scenario analysis, and what-if analysis, which correspond to the Pearl theory of causation: association, intervention, and counterfactual reasoning. Process analysis focuses on individual interactions in time and space, especially key thresholds and key attributes. Scenario analysis can provide comprehensive methods for experimental simulation analysis. Also, What-if analysis can compare models using different rule sets, such as the difference between Moore neighborhood and standard von Neumann neighborhood.

The public policy simulation represented by "Decision Theater" is a good example of this perspective [114]. By presenting the "if so, then" scenario, counterfactual reasoning can be made on the consequences of different policies. Its specific workflow is as follows: **1)** First, collect a large amount of historical data on the problems to be solved and organize them according to different dimensions; **2)** Second, the problem model is established by government decision-makers and experts, and then the abstract expert knowledge is transformed into an acceptable scene language using computational experiments; **3)** Third, the decision-making scenarios, in reality, are simulated and visualized to provide multiple policy scenarios and result previews for policy makers; **4)** Finally, in an agreed decision environment, various interest groups choose a satisfactory solution to policy issues. The utilization of water resources in Phoenix City is a typical application case [115].

*C. Q3: How to implement guiding intelligence?*

In complex systems, individual activities are increasingly complex, diverse, and differentiated, while intervention is simple, identical, and average centralized control. Considering the dynamics and uncertainties of the external environment, the intervention will lead to adaptive changes of some individuals according to the surrounding environment, which will lead to the failure of the initial optimization method. Therefore, the design of an intervention strategy is not a task that has been accomplished in one stroke. Instead, it needs to be constantly adjusted and optimized according to changes in the external environment. Computational experiments can build the bridge between the virtual world and the real world, eliminate the unfeasible strategies by counterfactual reasoning, and transfer the optimized strategies to the real space to guide the operation of the real system.

In the virtual world, extreme scene can exist and occur, which is rare in the real world. Based on the optimization theory of operations research on uncertain and multi-objective conditions, the intervention strategies are constantly trained and revised to obtain the expected optimization results. The aim is to intervene in the current development trajectory and running state from a future perspective, find out the possible adverse effects, conflicts, and potential dangers, and provide "reference", "prediction" and "guidance" for possible scenarios. Reinforcement learning is suitable for solving these problems. Its core task is to learn a strategic function, to maximize a long-term benefit and to establish a mapping relationship between observed values and output behaviors. The intervention strategy continues to evolve through machine learning methods in the loop feedback process with the test scenario.

Further, a collaborative dynamic feedback control system may be constituted between computational experiments and a real system. **1)** First, by analyzing the dependence of data and behavior, real-world problems are abstracted into the cognitive space. **2)** Second, a model is established in the cognitive space for computational experiments. The obtained results can dynamically guide or control strategy execution in the real world. **3)** Third, through the game process between the intervention strategy and the real world, the action space with the greatest difference is explored to reduce the modeling error of the artificial system and the execution error of the intervention strategy. **4)** Finally, the real-world execution results, in turn, continuously update the cognitive space model in the form of dynamic data input, forming a cyclic feedback learning process. The system iterates in this way of parallel execution until convergence. In the continuous integration of growing data, the training and simulation of artificial society will become more real and accurate. Through interactive feedback with the real world, it can obtain self-growth and self-optimization ability.

VI. CONCLUSION

In scientific research, induction is the discovery of patterns in empirical data, which is widely used in opinion surveys and macro-law analysis; deduction is to define a set of axioms and prove the conclusions that can be derived from these axioms. Compared with deduction and induction, the computational experiment method can be regarded as the third methodology of scientific research. Similar to deduction methods, computational experiments start by defining a clear set of assumptions; but unlike deduction methods, computational experiments do not prove theorems. Computational experiments can produce simulation data that can be analyzed by inductive methods; but unlike typical induction methods, simulation data comes from the emergence of rules operation rather than direct measurements of the real world. The purpose of induction is to discover patterns from data, and the purpose of deduction is to discover the results caused by hypotheses. Compared with them, computational experiments can verify our intuition through systematic deduction, trial & error, and estimation in the virtual world.

This paper summarizes the present situation and challenges of computational experiment methods, including the conceptual origin, research framework, application cases, and future challenges. The framework of computational experiments mainly consists of the modeling of artificial society, construction of experiment system, experimental design, experimental analysis, and experimental verification. The computational experiments' application cases can be summarized into three categories: thought experiment, mechanism exploration, and parallel optimization, involving political, economic, commercial, and epidemiological dimensions, etc. The technical challenges of computational experiments focus on how to combine with emerging smart technologies, including describing intelligence, predictive intelligence, and guiding intelligence.

Computational experiments provide new perspectives for the analysis and study of complex systems. Although relevant research has made great progress, there are still many problems and challenges in experimental design, experimental analysis, and experimental verification. For example, whether the computational experiment results match the known system laws (correctness); whether the representation of the computational model is elegant, including rules, syntactic structure, and similar features (conciseness); whether computational experiments can help transform the real world (effectiveness), etc.

In essence, computational experiments are designed to solve one or more research problems defined by a realistic reference system. Theoretically, the representation form of computational experiments is relatively simple, but the real system is more complex. If the abstract degree of the computational system is too high, it will be difficult to reflect the operation law of the real world. If the abstract degree of the computational system is too low, it will be too complex to build a model. A series of problems (lack of data, insufficient resources and time, and lack of knowledge) will be encountered. Therefore, the development of computational experiments need to seek a balance between the flexibility of modeling and the credibility of conclusion. In addition, computational experiments need to introduce more general and effective experimental mechanisms and analysis models to solve the problems of analysis, design, management, control, and synthesis of complex systems.

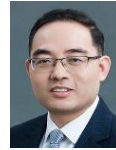
**Xiao Xue,** born in 1979. Professor at the School of Computer Software, College of Intelligence and Computing, Tianjin University. Also an adjunct professor at the School of Computer Science and Technology, Henan Polytechnic University. His main research interests include service computing, computational experiment, the Internet of things, etc.

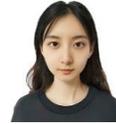
**Xiangning Yu**, born in 2000. Undergraduate student at the College of Intelligence and Computing, Tianjin University. Her current research interests include service computing and computational experiment.

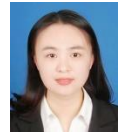
**Deyu Zhou,** born in 1995. PhD student in the School of Software, Shandong University, Jinan, China. Her current research interests include service computing and computational experiment.

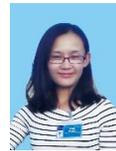
**Xiao Wang**, Associate Professor in the State Key Laboratory for Management and Control of Complex Systems, Institute of Automation, Chinese Academy of Sciences, and executive president of Qingdao Academy of Intelligent Industries. Her research interests include social network analysis, social transportation, cybermovement organizations, and multi-agent modeling.

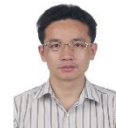
**Zhangbing Zhou,** born in 1975. Professor at China University of Geosciences, Beijing, China, and Adjunct Professor at TELECOM SudParis, Evry, France. He has authored over 100 referred papers. His research interests include process-aware information systems and sensor network middleware.

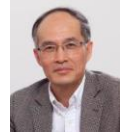
**Fei-Yue Wang**, State Specially Appointed Expert and Director of the State Key Laboratory for Management and Control of Complex Systems. His current research focuses on methods and applications for parallel systems, social computing, parallel intelligence, and knowledge automation.